\documentstyle[12pt]{article}
\newlength{\mytopmargin}
\newlength{\myleftmargin}
\begin{document}
\noindent
\begin{center}{  \Large\bf
 THE CALOGERO-SUTHERLAND MODEL AND \\[2mm]  POLYNOMIALS WITH
PRESCRIBED \\[4mm] SYMMETRY}
\end{center}
\vspace{5mm}

 \noindent
 \begin{center} T.H.~Baker\footnote{email: tbaker@maths.mu.oz.au} and
 P.J.~Forrester\footnote{email: matpjf@maths.mu.oz.au}\\[2mm]
 {\it Department of Mathematics, University of Melbourne, \\
  Parkville, Victoria 3052, Australia}
\end{center}
 \vspace{.5cm}

\begin{quote}
The Schr\"odinger operators with exchange terms for certain 
Calogero-Sutherland quantum many body systems have eigenfunctions 
which factor into the symmetric ground state and a multivariable 
polynomial. The polynomial can be chosen to have a prescribed 
symmetry (i.e.~be symmetric or antisymmetric) with respect to
the interchange of some specified variables. For four particular 
Calogero-Sutherland systems we construct an eigenoperator for these 
polynomials which separates the eigenvalues and establishes 
orthogonality. In two of the cases this involves identifying new 
operators which commute with the corresponding Schr\"odinger operators. 
In each case we express a particular class of the polynomials with 
prescribed symmetry in a factored form involving the corresponding 
symmetric polynomials.
\end{quote}
\date{} 

\section{Introduction}
\renewcommand{\theequation}{\thesection.\arabic{equation}}
The Schr\"odinger operator
\begin{equation}
H^{(C)} = - \sum_{j=1}^N {\partial^2 \over \partial x_j^2}
 +\beta (\beta /2 - 1){\left ( \pi \over L \right )^2 } \sum_{1 \le j < k \le N}
{1 \over \sin^2 \pi (x_j - x_k)/L}, \quad 0 \le x_j \le L
\label{HC}
\end{equation}
 describes quantum particles on a line of length $L$
 interacting through a $1/r^2$ pair potential with periodic boundary conditions,
or equivalently quantum particles on a circle of circumference length $L$
(hence the superscript $(C)$)
with the pair potential proportional to the inverse square of the chord length.
It is one of a number of quantum many body systems in one dimension
which are of the Calogero-Sutherland type, meaning that the ground state 
(i.e.~eigenstate with the smallest eigenvalue $E_0$) is
of the form $e^{-\beta W/2}$ with $W$ consisting of one and two body terms only.
Explicitly, for (\ref{HC}) we have
\begin{equation}
W = W^{(C)} := - \sum_{1 \le j < k \le N}
\log |e^{2 \pi i x_k/L} - e^{2 \pi i x_j/L}|.
\label{WC}
\end{equation}

In studying the integrability properties of (\ref{HC}) it is useful
\cite{poly92a} to generalize the Schr\"odinger operator to include the
exchange operator $M_{jk}$ for coordinates $x_j$ and $x_k$:
\begin{equation}
H^{(C,Ex)} = - \sum_{j=1}^N {\partial^2 \over \partial x_j^2}
 +\beta {\left ( \pi \over L \right )^2 } \sum_{1 \le j < k \le N}
{(\beta /2 - M_{jk}) \over \sin^2 \pi (x_j - x_k)/L},
\label{HCE}
\end{equation}
When acting on functions symmetric in $x_1,\dots, x_N$, (\ref{HCE}) reduces to
(\ref{HC}). Conjugation with the ground state of (\ref{HC}) gives the
transformed operator
\begin{eqnarray}\tilde{H}^{(C,Ex)}& := &
\Big ( {L \over 2 \pi} \Big )^2 e^{\beta W^{(C)}/2} (H^{(C,Ex)} - E_0^{(C)})
e^{-\beta W^{(C)}/2} \nonumber \\ & = &  \sum_{j=1}^N
\Big (z_j {\partial \over \partial z_j} \Big )^2 +  { N-1 \over \alpha}
\sum_{j=1}^N z_j {\partial \over \partial z_j}
+ {2 \over \alpha}  \sum_{1 \le j < k \le N}{z_j z_k \over z_j -
z_k}
\nonumber \\ & & 
\times \left[\Big ({\partial \over \partial z_j} -{\partial
\over 
\partial z_k}
\Big ) - {1 - M_{jk} \over z_j - z_k} \right]
\label{HCC}
\end{eqnarray}
where
\begin{equation}
z_j := e^{2 \pi i x_j / L}, \qquad \alpha := 2/\beta.
\end{equation}
This operator has non-symmetric eigenfunctions of the form
\begin{equation}
E_{\eta}(z,\alpha) = z^{\eta} + \sum_{\nu < \eta}
b_{\nu \eta} z^\eta.
\label{nj}
\end{equation}
($z^\eta$ will be referred to as the leading term),
where $\eta$ and $\nu$ are  $N$-tuples of non-negative integers and the
$b_{\nu \eta}$ are coefficients. To define the partial order $<$, 
let $\kappa$ be a
partition and $P$ be the (unique) permutation of minimal length such that 
\begin{equation}
 z^{\eta} := z_1^{\eta_1} \dots z_N^{\eta_N} =
z_{P(1)}^{\kappa_1} \dots z_{P(N)}^{\kappa_N},
\label{ETA}
\end{equation}
Equivalently, let $\kappa_i = \eta_{P(i)}$ and similarly define the partition
$\mu$ and permutation $Q$ such that $\mu_i = \nu_{Q(i)}$.
 The partial order $<$ is defined by the statement that $\nu < \eta$
if $\mu < \kappa$ (dominance ordering) or, in the cases $\mu = \kappa$, if the
first non-vanishing difference $P(j) - Q(j)$ is positive. An equivalent
specification in this later case is that the last nonvanishing difference of
$\eta - \nu$ is negative \cite{ug96a}. The eigenfunctions
$E_{\eta}(z,\alpha)$ are referred to as non-symmetric Jack
polynomials \cite{opdam95a,bern93a}. The eigenvalue of (\ref{HCC}) 
corresponding to the eigenfunction $E_{\eta}$ is given by \cite{kk95}
\begin{equation}
\epsilon_{\eta} = \sum_{j=1}^N \kappa_j^2 + {1 \over \alpha}
(N + 1 - 2j) \kappa_j
\label{KEV}
\end{equation}
and is thus independent of the permutation $P$ relating $\eta$ to $\kappa$. Note
 that this implies the linear combination
\begin{equation}
\sum_{P=1}^{N!} a_P E_{P^{-1}\kappa} (z,\alpha)
\label{SNS}
\end{equation}
is also an eigenfunction of  (\ref{HCC}).

The exchange operator generalization can also be applied to other 
Schr\"odinger  operators of the Calogero-Sutherland type, in particular when
the underlying root system is $A_N$ or $B_N$ and there is an external potential
or if the underlying root system is $BC_N$.
These operators are given by \cite{poly92a,yamam96a}
\begin{equation}
H^{(H,Ex)} :=  - \sum_{j=1}^N {\partial^2 \over \partial x_j^2}
+{\beta^2 \over 4} \sum_{j=1}^N x_j^2 +\beta  
\sum_{1 \le j < k \le N}
{(\beta /2 - M_{jk}) \over (x_j - x_k)^2}
\label{HHE}
\end{equation}
\begin{eqnarray}
H^{(L,Ex)} & := & - \sum_{j=1}^N {\partial^2 \over \partial x_j^2}
+{\beta^2 \over 4} \sum_{j=1}^N x_j^2 +
{\beta a' \over 2} \sum_{j=1}^N {\beta a' / 2 - S_j \over x_j^2} \nonumber \\
&&
+ \beta \sum_{1 \le j < k \le N} 
{\beta / 2 - M_{jk} \over (x_j - x_k)^2} + {\beta / 2 - S_j S_k M_{jk} \over (x_j
+ x_k)^2}
\label{HLE}
\end{eqnarray}
\begin{eqnarray}
H^{(J,Ex)} & := & - \sum_{j=1}^N {\partial^2 \over \partial \phi_j^2} +
{\beta a' \over 2} \sum_{j=1}^N {\beta a' / 2 - S_j \over \sin^2 \phi_j}
+ {\beta b' \over 2} \sum_{j=1}^N {\beta b' / 2 - S_j \over \cos^2 \phi_j}
\nonumber \\
& & +  \beta \sum_{1 \le j < k \le N} 
{\beta / 2 - M_{jk} \over \sin^2( \phi_j - \phi_k)} +
 {\beta / 2 - S_j S_k M_{jk} \over  \sin^2(\phi_j + \phi_k)}
\label{HJE}
\end{eqnarray}
respectively, where the operator $S_j$ replaces the coordinate $x_j$ ($\phi_j$)
by $-x_j$ $(-\phi_j)$ (the superscripts $(H)$, $(L)$ and $(J)$ stand for
Hermite,  Laguerre and Jacobi due to the relationship with these classical
polynomials in the case $N=1$ \cite{forr96a}). 

The symmetric ground state eigenfunctions of
(\ref{HHE}) and (\ref{HLE}) are of the form $e^{-\beta W}$ with $W$ given by
\begin{equation}
W^{(H)} := {1 \over 2} \sum_{j=1}^N x_j^2 - \sum_{1 \le j < k \le N}
\log |x_k - x_j|
\label{WH}
\end{equation}
\begin{equation}
 W^{(L)} := {1 \over 2} \sum_{j=1}^N x_j^2  -{a'\over 2} \sum_{j=1}^N \log
x^2 -  \sum_{1 \le j < k \le N} \log |x_k^2 - x_j^2|,
\label{WL}
\end{equation}
\begin{equation}
W^{(J)} :=  -{a'\over 2} \sum_{j=1}^N \log  \sin^2 \phi_j
 -{b'\over 2} \sum_{j=1}^N \log  \cos^2 \phi_j
- \sum_{1 \le j < k \le N}
\log | \sin^2 \phi_j -  \sin^2 \phi_k|.
\label{WJ}
\end{equation} 
Also, the transformation analogous to (\ref{HCC}) gives
\begin{eqnarray}\tilde{H}^{(H,Ex)} & := & 
- {2 \over \beta} e^{\beta W^{(H)}/2} (H^{(H,Ex)} - E_0^{(H)})
e^{-\beta W^{(H)}/2} \nonumber \\ &  = &  \sum_{j=1}^N
\Big ( {\partial^2 \over \partial y_j^2 } - 2 y_j {\partial
 \over \partial y_j } \Big ) + {2 \over \alpha} \sum_{j < k}
{1 \over y_j - y_k} \left[ \Big ( {\partial \over \partial y_j } -
 {\partial \over \partial y_k } \Big ) - {1 - M_{jk} \over y_j - y_k}
\right]
\label{HTHE}
\end{eqnarray}
\begin{eqnarray}\tilde{H}^{(L,Ex)} & := & 
- {1 \over 2 \beta} e^{\beta W^{(L)}/2} (H^{(L,Ex)} - E_0^{(L)})
e^{-\beta W^{(L)}/2} \nonumber \\ &  = &  \sum_{j=1}^N \Big (
y_j {\partial^2 \over \partial y_j^2 } + (a+1 - y_j) 
{\partial \over \partial y_j } \nonumber \\
&& + {1 \over \alpha} \sum_{j < k} {1 \over y_j - y_k}
\left[ 2 \Big ( y_j {\partial \over \partial y_j } - 
y_k  {\partial \over \partial y_k } \Big ) -
{y_j + y_k \over y_j - y_k} (1 -M_{jk}) \right]
\label{HTLE}
\end{eqnarray}
\begin{eqnarray}\tilde{H}^{(J,Ex)} & := & -{1 \over 4} 
 e^{\beta W^{(J)}/2} (H^{(J,Ex)} - E_0^{(J)})
e^{-\beta W^{(L)}/2} \nonumber \\ &  = &  \sum_{j=1}^N \Big (
z_j {\partial \over \partial z_j} \Big )^2 +
\Big ( (a + 1/2) {z_j + 1 \over z_j - 1} +
 (b + 1/2) {z_j - 1 \over z_j + 1} +{2(N-1) \over \alpha} 
\Big ) z_j {\partial \over
\partial z_j} \nonumber \\ &&
+{2 \over \alpha}  \sum_{1 \le j < k \le N}{z_j z_k \over z_j -z_k}
 \left[\Big ({\partial \over \partial z_j} -{\partial \over  \partial z_k}
\Big ) - {1 - M_{jk} \over z_j - z_k} \right] \nonumber \\ &&
+ {2 \over \alpha}  \sum_{1 \le j < k \le N}
{1 \over z_jz_k - 1} \left[\Big (z_j{\partial \over \partial z_j} -
z_k{\partial \over  \partial z_k}
\Big ) - {z_j z_k (1 - M_{jk}) \over z_j z_k - 1} \right]
\label{HTJE}
\end{eqnarray}
 where
\begin{equation}
a:= (a'\beta - 1) / 2,  \qquad b:= (b'\beta - 1) / 2.
\end{equation}
To obtain (\ref{HTHE}) we have made the change of variables $y_j = \sqrt{\beta
/ 2}\, x_j$, while to obtain (\ref{HTLE})  we have made the change of variables
$y_j = \beta x_j^2 / 2$ and imposed the restriction to eigenfunctions which are
even in $x_j$, and to obtain (\ref{HTJE}) we have used the variable
$z_j = e^{2i \phi_j}$ and imposed the restriction to eigenfunctions unchanged
by the mapping $z_j \mapsto 1/z_j$.

Our objective is to initiate a study into properties of the polynomial
eigenfunctions of the operators (\ref{HTHE})-(\ref{HTJE}), and to supplement
some of the results of \cite{forr96c,forr96b} on the polynomial eigenfunctions
of (\ref{HCC}) with a prescribed symmetry (i.e.~eigenfunctions which are
either symmetric or anti-symmetric with respect to the interchange of
specified variables). In Section 2 we consider (\ref{HCC}). We revise the
construction of an eigenoperator for the symmetric polynomial eigenfunctions
(the Jack polynomials) which separates the eigenvalues, and how it can be used
to establish orthogonality. This construction is then generalized to provide an
eigenoperator for the Jack polynomials with prescribed symmetry, which is
used to establish orthogonality.

In Sections 3 and 4 we introduce non-symmetric generalized Hermite and
Laguerre polynomials as eigenfunctions of (\ref{HTHE}) and (\ref{HTLE})
respectively. Exponential operator formulas are given relating these
polynomials to the non-symmetric Jack polynomials. New commuting operators are
identified which are used to prove the orthogonality of these polynomials.
Generalized Hermite and Laguerre polynomials with prescribed symmetry
are defined, and the commuting operators are used to define an eigenoperator
which separates the eigenvalues and establishes orthogonality.

In Section 5 we begin by revising known commuting operators which decompose
the operator (\ref{HTJE}). Non-symmetric generalized Jacobi polynomials,
and Jacobi polynomials with prescribed symmetry are defined as eigenfunctions
of these operators, and orthogonality is established. 

The final subsection of Section 2-5 is devoted to establishing a formula
expressing a particular class of the polynomials with prescribed symmetry 
in a factored form involving the corresponding symmetric polynomials. 

\section{Eigenfunctions of $\tilde{H}^{(C,Ex)}$}
\setcounter{equation}{0}
\renewcommand{\theequation}{\thesection.\arabic{equation}}
\subsection{Revision}
The operator $\tilde{H}^{(C,Ex)}$ allows a factorization in terms of so-called
Cherednik operators $\hat{D}_j$ \cite{cher91a}. These operators 
are given in terms of the Dunkl operator
\begin{equation}
T_j := {\partial \over  \partial z_j} + {1 \over \alpha}  \sum_{k=1 \atop
k \ne j} {1 \over z_j - z_k} (1 - M_{jk})
\label{DU}
\end{equation}
for the root system $A_{N-1}$ by
\begin{eqnarray}
\hat{D}_j & := & z_j T_j + {1 \over \alpha} \sum_{k=1}^{j-1} M_{jk}
\nonumber \\ & = & z_j {\partial \over  \partial z_j} +
{1 \over \alpha} \Big ( \sum_{l < j} {z_l \over z_j - z_l} (1 - M_{lj})
+ \sum_{l > j} {z_j \over z_j - z_l} (1 - M_{lj}) \Big )
+ { (j-1) \over \alpha}.
\label{CO}
\end{eqnarray}
They mutually commute so that
\begin{equation}
[\hat{D}_j, \hat{D}_k] = 0.
\label{DOC1}
\end{equation}
This can be checked from the fact that the Dunkl operators commute:
\begin{equation}
[T_j,T_k]=0.
\label{DOC}
\end{equation}

The non-symmetric Jack polynomials are simultaneous eigenfunctions of the
$\hat{D}_j$ for each $j=1,\dots,N$, and the corresponding eigenvalues are
\begin{equation}
e_{j,\eta} = \eta_j + {1 \over \alpha}
\Big ( -  \sum_{l < j} h(\eta_l - \eta_j) +  \sum_{l > j} h(\eta_j - \eta_l)
\Big ) + {(j-1) \over \alpha}
\label{EV}
\end{equation}
with
\begin{equation}
h(x) = \bigg \{ \begin{array}{l} 1,\qquad x>0 \\ 0, \qquad {\rm otherwise.}
\end{array}
\label{HX}
\end{equation}
We remark that some authors \cite{sahi96a,knop96c} define non-symmetric Jack
polynomials as eigenfunctions of a variant of the Cherednik operators
(\ref{CO}):
\begin{eqnarray}
\xi_j & := & \alpha y_j T_j + \sum_{k=j+1}^N M_{jk} + (1-N) \nonumber \\
& = &
\alpha \left[ y_j {\partial \over  \partial y_j} +
{1 \over \alpha} \Big ( \sum_{l < j} {y_j \over y_j - y_l} (1 - M_{lj})
+ \sum_{l > j} {y_l \over y_j - y_l} (1 - M_{lj}) \Big )
- { (j-1) \over \alpha} \right].
\label{COV}
\end{eqnarray}
(the choice of notation $y_1,\dots,y_N$ for the coordinates here is for later
convenience).
The operators $\xi_j$ have the same polynomial eigenfunctions as $\hat{D}_j$
except that $z_j$ is replaced by $y_{N+1 - j}$ $(j=1,\dots,N)$ and $\eta_j$ is
replaced by $\eta_{N+1-j}$. Following
\cite{knop96c,sahi96a} the corresponding eigenvalues are to be 
denoted $\bar{\eta}_j$ and are explicitly given by
\begin{equation}
\bar{\eta}_j = \alpha \eta_j - \Big ( \sum_{l < j} h(\eta_l + 1 - \eta_j)
+ \sum_{l > j} h(\eta_l - \eta_j) \Big ).
\label{EVCV}
\end{equation}

Returning to the relationship to $\tilde{H}^{(C,Ex)}$, a direct calculation
shows \cite{bern93a}
\begin{equation}
\tilde{H}^{(C,Ex)} = \sum_{j=1}^N \left( \hat{D}_j - {N-1 \over 2 \alpha}
\right)^2 - \Big ( {L \over 2 \pi} \Big )^2 E_0^{(C)}.
\label{FAC}
\end{equation}
Let us revise \cite{bern93a} how this decomposition 
can be used to prove that the
symmetric polynomial eigenfunctions of (\ref{HCC}) (i.e.~the symmetric Jack
polynomials) are orthogonal with respect to the inner product
\begin{equation}
\langle f | g \rangle^{(C)} := \int_0^L dx_1 \dots \int_0^L dx_N \,
\prod_{1 \le j < k \le N} |z_k - z_j|^{2 /\alpha} f(z_1^*,\dots,z_N^*)
g(z_1,\dots,z_N),
\label{IC}
\end{equation}
where $ z_j = e^{2 \pi i x_j /L}$ and $^*$ denotes complex conjugate. First we
check directly from the definitions (\ref{CO}) and (\ref{IC}) that
\begin{equation}
\langle f |  \hat{D}_j g \rangle^{(C)} = \langle  \hat{D}_j f | g \rangle^{(C)}
\label{SA}
\end{equation}
which says that the Cherednik operators are self-adjoint with respect to
the inner product  (\ref{IC}). In performing this check we use the facts
that
$$
\prod_{1 \le j < k \le N} |z_k - z_j|^{2 /\alpha} = \psi_0^* \psi_0, \qquad
{\rm where} \quad  \psi_0 = \prod_{j=1}^N z_j^{-(N-1)/\alpha}
\prod_{1 \le j < k \le N} (z_k - z_j)^{1/\alpha}
$$
and
$$
\psi_0 \hat{D}_j \psi_0^{-1} = 
 z_j {\partial \over  \partial z_j} -
{1 \over \alpha} \left( \sum_{l < j} {z_l \over z_j - z_l}  M_{lj}
+ \sum_{l > j} {z_j \over z_j - z_l}  M_{lj} \right)
+ { (N-1) \over 2\alpha}.
$$
Note that $|\psi_0|^2 = e^{-\beta W^{(C)}}$, where $W^{(C)}$ is given by 
(\ref{WC}), and is thus the square of the ground state wave function for
(\ref{HC}).
Next, we compare the eigenvalues
$\{ e_{j,\eta} \}$ to $\{ e_{j,\eta'} \}$, where
$\eta'$ is obtained from the $N$-tuple $\eta$ by interchanging $\eta_i$
and $\eta_{i'}$.

\vspace{.2cm}
\noindent {\bf Lemma 2.1} \quad We have
$$
 e_{i,\eta'}= e_{i',\eta}, \quad
 e_{i',\eta'}= e_{i,\eta} \quad {\rm and} \quad
 e_{j,\eta'}= e_{j,\eta}, \quad (j \ne i',i).
$$

\vspace{.2cm}
\noindent {\bf Proof} \quad These equations are verified directly from
(\ref{EV}).

\vspace{.2cm}
\noindent {\bf Remark} \quad The result analogous to Lemma 2.1 applies for the
eigenvalues (\ref{EVCV}).

\vspace{.2cm}
{}From Lemma 2.1 we see that $\{ e_{j,\eta} \}_{j=1,\dots,N}$ with
$\eta = P^{-1} \kappa$ is independent of the permutation $P$. Choosing the
permutation $P(j) = N + 1 -j$ $(j=1,\dots,N)$ shows that
$\{ e_{j,\eta} \}_{j=1,\dots,N} = \{ \kappa_{N+1-j}
+ (j-1)/\alpha \}_{j=1,\dots,N}$. This allows an eigenoperator of the
$E_{P^{-1}\kappa}$ to be constructed for which the eigenvalues are
independent of $P$:
\begin{eqnarray}\lefteqn{
\Big ( 1 + u(\hat{D}_1 - (N-1)/2\alpha) \Big ) \dots
\Big ( 1 + u(\hat{D}_N - (N-1)/2\alpha) \Big )
E_{P^{-1}\kappa}}\nonumber 
\\ && \hspace*{2cm} 
= \prod_{j=1}^N \Big ( 1 + u(\kappa_j + (N+1 - 2j)/2\alpha) \Big ) 
 E_{P^{-1}\kappa}
\label{EVC}
\end{eqnarray}
(the constants $ - (N-1)/2\alpha$ are not essential and could have been
omitted). Note that 
\begin{equation}
\kappa_1 -{1 \over \alpha} > \kappa_2 - {2 \over \alpha} > \dots >
\kappa_N - {N \over \alpha}
\label{OR}
\end{equation}
so the eigenvalues for different partitions $\kappa$ are distinct.

Consider now the symmetric Jack polynomials $J_\kappa^{(\alpha)}$. They can be
characterized (up to normalization, which for definiteness we will take to be
that adopted by Stanley \cite{stan89a}) as the polynomial eigenfunctions of
(\ref{HCC}) with leading term
$m_\kappa$. From the fact that the non-symmetric Jack polynomials
$E_{P^{-1}\kappa}$ are simultaneous eigenfunctions of all the $\hat{D}_j$ with
leading term
$z^{P^{-1}\kappa}$ and the triangular structure (\ref{nj})
we must have
\begin{equation}
J_\kappa^{(\alpha)}(z) = \sum_{P} a_{P^{-1}\kappa} E_{P^{-1}\kappa}(z,\alpha)
\label{JNSJ}
\end{equation}
for some coefficients $a_{P^{-1}\kappa}$ (these 
coefficients are given explicitly
in \cite{sahi96a}). It follows that the symmetric Jack polynomial satisfies the
eigenvalue equation (\ref{EVC}). Since by (\ref{SA}) the operator in
(\ref{EVC}) is self-adjoint with respect to the inner product (\ref{IC})
and by (\ref{OR}) the eigenvalues are distinct, this implies that the
symmetric Jack polynomials are orthogonal with respect to (\ref{IC}).

With respect to the eigenvalue equation (\ref{EVC}) with $E_{P^{-1}\kappa}$
replaced by $J_\kappa^{(\alpha)}$, we remark that Macdonald \cite{mac} has
constructed an operator $D_N(X;\alpha)$ such that
\begin{equation}
D_N(X;\alpha) \,J_\kappa^{(\alpha)} = \prod_{i=1}^N (X + N - i + \alpha
\kappa_i) \, J_\kappa^{(\alpha)}.
\end{equation}
Since $\{ J_\kappa^{(\alpha)} \}$ is a basis for symmetric functions, it follows
by comparison with (\ref{EVC}) that when acting on symmetric functions
\cite{noumi96a,noumi96b}
\begin{equation}
\prod_{j=1}^N (X + \alpha \hat{D}_j) = D_N(X;\alpha).
\end{equation}

Another way of establishing the orthogonality of the symmetric Jack polynomials
is to use the expansion (2.14) together with the fact that the non-symmetric
Jack polynomials form an orthogonal set with respect to (\ref{IC}). This later
fact can be established by first noting that
\begin{equation}
\prod_{l=1}^N(1 + u_l \hat{D}{_l})
\label{PONS}
\end{equation}
is an eigenoperator of each $E_\eta$ which separates the eigenvalues. The 
result now follows after using the fact that (\ref{PONS}) is self-adjoint.

\subsection{Jack polynomials with prescribed symmetry}
As noted above, the fact that the non-symmetric Jack polynomials $E_\eta$
are simultaneous eigenfunctions of $\hat{D}_1, \dots,\hat{D}_N$ implies that
the $E_\eta$ are eigenfunctions of (\ref{HCC}). Since (\ref{HCC}) is symmetric
in $z_1,\dots,z_N$ it follows that $E_\eta$ with the variables 
$z_1,\dots,z_N$ permuted is also an eigenfunction of (\ref{HCC}) with the same
eigenvalue. Thus, for any permutation $P$, since the leading order term of 
$J_\kappa^{(\alpha)}(z)$ is proportional to the monomial symmetric function
(i.e.~the symmetrization of $z^{P^{-1}\kappa}$) we must have
\begin{equation}
J_\kappa^{(\alpha)}(z) = A_{P^{-1} \kappa}\,{\rm Sym} \, \Big (
E_{P^{-1}\kappa}(z,\alpha) \Big ).
\label{VK}
\end{equation}
For the case $P^{-1}\kappa=\kappa$, ${\rm Sym} \, \Big (
E_{P^{-1}\kappa}(z,\alpha) \Big )$ has leading term $m_{\kappa}$, so
$A_{P^{-1} \kappa} = v_{\kappa \kappa}$ where $ v_{\kappa \kappa}$ is
defined and given explicitly in ref.~\cite{stan89a}.
Eigenfunctions can be constructed in an analogous way which are symmetric with
respect to the interchange of certain sets of variables and antisymmetric with
respect to the interchange of other sets of variables. We will refer to such
polynomials as having a prescribed symmetry.
To facilitate a discussion of this situation, let us rewrite the coordinates
$\{z_j \}_{j=1,\dots,N}$ as
$$
\Big (\bigcup_{\alpha = 1}^q \{ w_j^{(\alpha)} \}_{j=1,\dots,N_\alpha^{(w)}} \Big
)
\Big (\bigcup_{\gamma = 1}^p \{ z_j^{(\gamma)} \}_{j=1,\dots,N_\gamma^{(z)}}
 \Big )
$$
taken in order so that $w_1^{(1)} = z_1, \dots, z_{N_p}^{(p)} = z_N$ and
$
N = \sum_{\mu = 1}^q N_\mu^{(w)} + \sum_{\gamma=1}^p N_\gamma^{(z)}.
$
We seek polynomial eigenfunctions of (\ref{HCC}), $S_{P^{-1}\kappa}(z,\alpha)$
say, which are symmetric in
$ \{ w_j^{(\mu)} \}_{j=1,\dots,N_\mu^{(w)}}$ and antisymmetric in 
$\{ z_j^{(\gamma)} \}_{j=1,\dots,N_\gamma^{(z)}}$. We have
\begin{equation}
S_{P^{-1}\kappa}(z,\alpha) = {\cal O} \Big ( E_{P^{-1}\kappa}(z,\alpha) \Big )
\label{DEFS}
\end{equation}
where ${\cal O}$ denotes the operation of symmetrization in $\{ w_j^{(1)}
\}_{j=1,\dots,N_1^{(w)}}$, antisymmetrization in
$\{ z_j^{(\gamma)} \}_{j=1,\dots,N_\gamma^{(z)}}$ and normalization
such that the coefficient of $z^{P^{-1}\kappa}$ is unity. 
Due to the operation
${\cal O }$ the label $P^{-1}\kappa$ in $S_{P^{-1}\kappa}$ can be replaced by
$q+p$ partitions $(\rho, \mu) := (\rho^{(1)}, \dots \rho^{(q)}, \mu^{(1)},
\dots, \mu^{(p)}$ where $\rho^{(\mu)}$ consists of $N_\alpha^{(w)}$ parts
$(\alpha =1,\dots,q)$ and $\mu^{(\gamma)}$ consists of $N_\gamma^{(z)}$ parts.
For the $N$-tuple $\eta = P^{-1}\kappa$ any 
rearrangements of
\begin{equation}
\{ \eta_j \}_{j=1,\dots,N_1^{(w)}}, \,
\{ \eta_{N_1^{(w)}+j} \}_{j=1,\dots,N_2^{(w)}}, \dots,
\{ \eta_{N^{(w)}+\sum_{\gamma=1}^{p-1}N_\gamma^{(z)}+j}
\}_{j=1,\dots,N_p^{(z)}},
\label{REA}
\end{equation}
where $N^{(w)} := \sum_{\mu = 1}^q  N_\mu^{(w)}$, 
give the same partitions $(\rho, \mu)$ and thus the same polynomial with
prescribed symmetry.

A feature of the polynomials $E_\eta(z,\alpha)$ is that if $\eta_i = \eta_{i+1}$
then $E_\eta$ is symmetric in $z_i$ and $z_{i+1}$ (see (\ref{swap}) below). It
follows that if two parts of any $\mu^{(\gamma)}$ are equal the polynomial
$S_{(\rho,
\mu)}$ vanishes identically due to the antisymmetrization procedure in its
construction. Thus each $\mu^{(\gamma)}$ must be restricted to distinct
parts.

Now we know from \cite[Lemma 2.4 and Proposition 4.3]{knop96c} that with
$s_i := M_{i \, i+1}$ and $\delta_i := \bar{\eta}_i - \bar{\eta}_{i+1}$,
\begin{equation}
s_i E_\eta = \left \{ \begin{array}{ll}
{1 \over \delta_i}  E_\eta + (1 - {1 \over \delta_i^2}) E_{s_i \eta},
&\quad \eta_i > \eta_{i+1} \\
E_\eta, &\quad  \eta_i = \eta_{i+1} \\
{1 \over \delta_i}  E_\eta +  E_{s_i \eta},
&\quad \eta_i < \eta_{i+1} \end{array} \right.
\label{swap} 
\end{equation}
(here $E_\eta$ refers to the eigenfunctions of (\ref{COV}), which as noted below
(\ref{COV}) are related to the non-symmetric Jacks defined as eigenfunctions of
(\ref{CO}) by relabelling).
Also, each permutation can be written as a product of the 
elementary transpositions $s_i$. Therefore, we
conclude that (\ref{DEFS}) can be rewritten as
\begin{equation}
S_{P^{-1}\kappa}(z,\alpha) = \sum_{\rm rearrangements} b_{Q^{-1}\kappa}
E_{Q^{-1}\kappa}(z,\alpha)
\label{SSUM}
\end{equation}
where the sum is over rearrangements $Q^{-1}\kappa$ of $P^{-1}\kappa$ obtained by
permuting within the sets (\ref{REA}).

Now two distinct sequences of partitions $(\rho, \mu)$ and $(\hat{\rho},
\hat{\mu})$ as defined below (\ref{DEFS}) cannot have any rearrangements of
(\ref{REA}) in common, as they wouldn't then be distinct. Hence the expansion
(\ref{SSUM}) for
$S_{(\rho, \mu)}$ and $S_{(\hat{\rho}, \hat{\mu})}$ does not contain any common
$E_\eta$. It follows immediately from the orthogonality of $\{E_\eta \}$ with
respect to (\ref{IC}) that $\{S_{(\rho, \mu)} \}$ are also orthogonal with
respect to (\ref{IC}).

An alternative way to deduce the orthogonality is to note that the operator
\begin{equation}
\prod_{\mu = 1}^q \prod_{j=1}^{N_\mu^{(w)}} \Big ( 1 + u_\mu
\hat{D}_{\sum_{l=1}^{\mu - 1}  N_l^{(w)} + j} \Big )
\prod_{\gamma = 1}^p \prod_{j=1}^{N_\gamma^{(z)}} \Big ( 1 + v_\gamma
\hat{D}_{N^{(w)}+\sum_{l=1}^{\gamma - 1}  N_l^{(z)} + j} \Big )
\label{PCO}
\end{equation}
is an eigenoperator of $\{ S_{(\rho, \mu)} \}$.
To see this, note that this is an eigenoperator of the non-symmetric Jacks, and
from Lemma 2.1 the corresponding eigenvalue is independent of the particular
rearrangements (\ref{REA}). Furthermore,
 the eigenvalues of
(\ref{PCO}) corresponding to  $S_{(\rho, \mu)}(z,\alpha)$
and $S_{(\hat{\rho}, \hat{\mu)}}(z,\alpha)$ are distinct  whenever 
$\rho \mu$ and $\hat{\rho} \hat{\mu}$ are distinct. Since, by
(\ref{SA}), (\ref{PCO}) is self-adjoint with respect to (\ref{IC}), the
fact that the eigenvalues are distinct 
implies orthogonality of these
functions with respect to the inner product  (\ref{IC}).

 In the case $q=0$, $p=2$, $|\mu^{(1)}| + |\mu^{(2)}| =1,2$
or 3, explicit formulas for $\tilde{S}_{\mu^{(1)}\mu^{(2)}}(z,\alpha)$ have
been given in ref.~\cite{yamam96b}, where the $\tilde{S}$ are eigenfunctions
of (\ref{HCC}) which are symmetric in $\{w_j^{(1)}\}$ and $\{w_j^{(2)}\}$.
However, in general $\tilde{S}$ does not correspond to $S$ as 
 $\tilde{S}$ does not satisfy (\ref{DEFS}), and the $\tilde{S}$ are not
orthogonal. 

\subsection{Some special Jack polynomials with prescribed symmetry}
In some previous works \cite{forr96c,forr96b} we have 
conjectured  a formula for
certain Jack polynomials with prescribed symmetry in terms of difference
products and the symmetric Jack polynomial. In the present notation the
conjecture in \cite{forr96c,forr96b} applies to 
$S_{(\rho,\kappa)}(z,\alpha)$ with $q=1$ and
\begin{equation}
(\rho,\kappa) = (\rho_1, \rho_2, \dots, \rho_{N_0}, N_1 - 1,N_1-2, \dots,
1,N_2-1, N_2-2, \dots ,1,N_p-1,N_p-2, \dots, 1)
\label{PK}
\end{equation}
(here we have written $N_1^{(w)} =: N_0$, $N_l^{(z)} =: N_l$ to be consistent
with refs.~\cite{forr96c,forr96b} and it is assumed 
$\rho_1 \ge \rho_2 \ge \cdots \ge \rho_{N_1} \ge 0$). 
The conjecture states that for $(\rho,\mu)$ given by (\ref{PK})
\begin{equation}
S_{(\rho,\mu)}(z,\alpha) = A_{(\rho,\mu)}
\prod_{\gamma=1}^p \prod_{1 \le j < k \le N_\gamma} (z_k^{(\gamma)} -
z_j^{(\gamma)}) J_{\rho}^{(p + \alpha)}(w_1,\dots,w_{N_0}),
\label{CON}
\end{equation}
where $ A_{\rho\mu}$ is some normalization, provided
\begin{equation}
\rho_1 \le {\rm min} (N_1,\dots,N_p)
\label{IN}
\end{equation}
(A stronger conjecture was also given in  \cite{forr96c,forr96b} which replaces
$\rho_1$ in (\ref{PK}) by $\rho_1-1$, however we do not consider that extension
here.)

This conjecture can be verified directly by showing that the r.h.s.~of
(\ref{CON}) is an eigenfunction of (\ref{HCC}) (an abbreviated version of the
required calculation was given in \cite{forr96c}, however the working there 
is incomplete and an erroneous conclusion was drawn). We begin by rewriting
the variables $z_1,\dots,z_N$ in (\ref{HCC}) as $\{w_j\}_{j=1,\dots,N_0}$
and $\{z_j^{(\gamma)} \}_{j=1,\dots,N_\gamma}$ $(\gamma = 1,\dots, p)$.
In terms of these variables, when acting on functions symmetric in
$\{w_j\}$ and anti-symmetric in $\{z_j^{(\gamma)}\}$ we have
\begin{equation}
\tilde{H}^{(C,Ex)} = \tilde{H}^{(C,w)} + \tilde{H}^{(C,z)} + \tilde{H}^{(C,wz)}
\label{DEC}
\end{equation}
where
\begin{equation}
\tilde{H}^{(C,w)} = 
  \sum_{j=1}^{N_0}
\Big (w_j {\partial \over \partial w_j} \Big )^2 +  { N-1 \over \alpha}
\sum_{j=1}^{N_0} w_j {\partial \over \partial w_j}
+ {2 \over \alpha}  \sum_{ j \ne k }{w_j w_k \over w_j -
w_k}{\partial \over \partial w_j}
\end{equation}
\begin{eqnarray}
\tilde{H}^{(C,z)} & = &
 \sum_{\gamma=1}^p \sum_{j=1}^{N_\gamma}
\Big (z_j^{(\gamma)} {\partial \over \partial z_j^{(\gamma)} } 
\Big )^2 +  { N-1 \over \alpha} \sum_{\gamma=1}^p \sum_{j=1}^{N_\gamma}
 z_j^{(\gamma)} {\partial \over \partial z_j^{(\gamma)}}
\nonumber \\ &&
+ {2 \over \alpha} \sum_{\gamma < \nu} \sum_{j=1}^{N_\gamma}
\sum_{k=1}^{N_\nu} {z_j^{(\gamma)} z_k^{(\nu)} \over z_j^{(\gamma)} -
 z_k^{(\nu)}}  \left[\Big ({\partial \over \partial z_j^{(\gamma)}} -{\partial
\over 
\partial z_k^{(\nu)}}
\Big ) - {1 - M(z_j^{(\gamma)},z_k^{(\nu)}) \over z_j^{(\gamma)} - z_k^{(\nu)}}
\right] \nonumber \\
&& + {2 \over \alpha}  \sum_{\gamma=1}^p \sum_{j<k}
{z_j^{(\gamma)} z_k^{(\gamma)} \over z_j^{(\gamma)} -
 z_k^{(\gamma)}}  \left[\Big ({\partial \over \partial z_j^{(\gamma)}} -
 {\partial \over \partial z_k^{(\gamma)}}
\Big ) - {2 \over z_j^{(\gamma)} - z_k^{(\gamma)}}
\right]
\end{eqnarray}
\begin{eqnarray}
\tilde{H}^{(C,wz)} = {2 \over \alpha} \sum_{\gamma = 1}^p  \sum_{j=1}^{N_\gamma}
\sum_{k=1}^{N_0}  {z_j^{(\gamma)} w_k \over z_j^{(\gamma)} -
 w_k}  \Big [\Big ({\partial \over \partial z_j^{(\gamma)}} -{\partial
\over 
\partial w_k}\Big ) - {1 - M(z_j^{(\gamma)},w_k) \over z_j^{(\gamma)} -
w_k} \Big ]
\label{DEFHCWZ}
\end{eqnarray}

Here we have used the notation $M(x,y)$ to denote the operator which exchanges
the coordinates $x$ and $y$.
We seek the action of these operators on
\begin{equation}
\prod_{\gamma = 1}^p \Delta (z^{(\gamma)}) \prod_{l=1}^{N_0} w_l^{\kappa_l},
\qquad \Delta (z^{(\gamma)}) := \prod_{1 \le j < k \le N_\gamma}(
z_k^{(\gamma)} - z_j^{(\gamma)}).
\label{DEL}
\end{equation}
For this purpose we require the following result \cite{kk95}, which can be
verified directly.

\vspace{.2cm}
\noindent
{\bf Lemma 2.2} \quad Let
$$
A(y_j,y_k) := {y_j y_k \over y_j - y_k} 
 \left[\Big ({\partial \over \partial y_j} -{\partial
\over 
\partial y_k}\Big ) - {1 - M(y_j,y_k) \over y_j -
y_k} \right]
$$
For $\lambda_j \ge \lambda_k$ we have
\begin{equation}
A(y_j,y_k) \, y_j^{\lambda_j} y_k^{\lambda_k} = -\lambda_k  y_j^{\lambda_j}
y_k^{\lambda_k} + \left \{ \begin{array}{l}\displaystyle{
\sum_{l=1}^{\lambda_j - \lambda_k-1}} (\lambda_j - \lambda_k - l)
y_j^{\lambda_j-l} y_k^{\lambda_k+l}, \quad \lambda_j - \lambda_k \ge 2 \\
0, \quad {\rm otherwise.} \end{array} \right.
\label{AA}
\end{equation}

\vspace{.2cm}
This lemma will first be used to determine the action of $H^{(C,z)}$ on 
(\ref{DEL}).

\vspace{.2cm}
\noindent
{\bf Lemma 2.3} \quad Let $P_\gamma$ be a permutation of
$\{1,2,\dots,N_\gamma\}$, and let
\begin{equation}
\label{PHI}
\Phi(z) := \prod_{\gamma=1}^p \prod_{j=1}^{N_\gamma}
 (z_j^{(\gamma)})^{P_\gamma(l) - 1}.
\end{equation}
We have
\begin{equation}
\tilde{H}^{(C,z)} \Phi(z) = \delta^{(C)}  \Phi(z) + \Omega (z).
\label{L1}
\end{equation}
where $\delta^{(C)}$ is independent of the permutations 
$P_\gamma$ and $\Omega(z)$ is a polynomial such that the exponent of
each monomial has at least one repeated part. Hence
\begin{equation}
\tilde{H}^{(z)}\,  \prod_{\gamma=1}^p \Delta (z^{(\gamma)}) =
 \delta^{(C)}  \prod_{\gamma=1}^p \Delta (z^{(\gamma)}).
\label{L2}
\end{equation}

\vspace{.2cm}
\noindent
{\bf Proof} \quad The fact that $\delta^{(C)}$ is independent of
 the permutations 
$P_\gamma$ follows from the eigenvalue (\ref{KEV}) being independent of the
permutation, while each monomial having at least one repeated part is a
consequence of the fact that the exponents in $\Phi(z)$ for each set of
variables $\{z_j\}_{j=1,\dots,N_\gamma}$ consists of the consecutive
integers $0,1,\dots,N_\gamma-1$, and the action of the operator
$A(z_j^{(\gamma)},z_k^{(\mu)})$ noted in Lemma 2.2. Anti-symmetrizing
both sides of (\ref{L1})  in the variables $\{z_j\}_{j=1,\dots,N_\gamma}$
($\gamma = 1,\dots,p$) the polynomial $\Omega (z)$ therefore gives zero
contribution, and the result (\ref{L2}) follows from the Vandermonde
determinant formula
\begin{equation}
\sum_{P=1}^{N!} \epsilon(P) \prod_{l=1}^N z_l^{P(l) - 1} =
\Delta (z),
\end{equation}
where $\epsilon(P)$ denotes the partity of the permutation $P$.

\vspace{.2cm}
The crucial point in establishing (\ref{CON}) is the action of
$H^{(C,wz)}$ on (\ref{DEL}).

\vspace{.2cm}
\noindent
{\bf Lemma 2.4} \quad Let $F(w,z) = m_\kappa (w)  \prod_{\gamma=1}^p \Delta
(z^{(\gamma)})$, where $\kappa$ is a partition consisting of $N_0$ parts
with the largest part $\kappa_1$ restricted by $\kappa_1 \le
{\rm min}(N_1,\dots,N_\gamma)$ and $ m_\kappa (w)$ 
denotes the monomial symmetric function with exponent
$\kappa$.
We have 
\begin{equation}
\tilde{H}^{(C,wz)}F(w,z) = \left({p \over \alpha} \sum_{j=1}^{N_0} \Big ( w_j
{\partial
\over
\partial w_j} \Big )^2 - {2 \over \alpha} (N^{(z)} - p/2) \sum_{j=1}^{N_0}
 w_j {\partial \over \partial
w_j} \right) F(w,z). \label{L3}
\end{equation}

\vspace{.2cm}
\noindent
{\bf Proof} \quad 
Consider first the action of $A(z_j^{(\gamma)},w_k)$ on $\Delta
(z^{(\gamma)}) w_k^\lambda$. 
Expanding $ \Delta
(z^{(\gamma)})$ into terms of the form (\ref{PHI}) we see from the argument of
the proof of Lemma 2.3 that for $0 \le \lambda \le N_{\gamma}$ only the first
term on the r.h.s. of (\ref{AA}) for the action of $A(z_j^{(\gamma)},w_k)$
on $z_j^{(\gamma)\lambda_j} w_k^\lambda$ contributes, and thus
$$
\sum_{j=1}^{N_\gamma} A(z_j^{(\gamma)},w_k) \Delta(z^{(\gamma)}) w_k^\lambda
= \Big ( - \sum_{j=1}^{N_\gamma} {\rm min} (\lambda, N_\gamma - j) \Big )
 \Delta(z^{(\gamma)}) w_k^\lambda.
$$
But for $0 \le \lambda \le N_\gamma$ a straightforward calculation gives
$$
-  \sum_{j=1}^{N_\gamma} {\rm min} (\lambda, N_\gamma - j) =
{1 \over 2} \lambda^2 - (N_\gamma - {1 \over 2}) \lambda.
$$
Thus
$$
\sum_{k=1}^{N_0}
 \sum_{j=1}^{N_\gamma} A(z_j^{(\gamma)},w_k) \Delta(z^{(\gamma)})
w_1^{\kappa_{Q(1)}} \dots w_{N_0}^{\kappa_{Q(N_0)}} =
\Big ( {1 \over 2} |\kappa^2| - (N_\gamma - {1 \over 2})|\kappa| \Big )
w_1^{\kappa_{Q(1)}}  \dots w_{N_0}^{\kappa_{Q(N_0)}},
$$
where $|\kappa| := \sum_{j=1}^{N_0} \kappa_j$, $|\kappa^2| := \sum_{j=1}^{N_0}
\kappa_j^2$, independent of the permutation
$Q$. Summing over $\gamma$ and comparison with the definition of
$\tilde{H}^{(C,wz)}$ shows that
$$
\tilde{H}^{(C,wz)} F(w,z)
= {2 \over \alpha} \Big ( {p \over 2} |\kappa^2| - (N^{(z)} - {p \over 2}
|\kappa|
\Big ) F(w,z).
$$
This equation remains valid with $H^{(C,wz)}$ replaced by the operator on the
r.h.s.~of (\ref{L3}), thus verifying the validity of (\ref{L3}).

\vspace{.2cm}
 Substituting the results of Lemmas 2.3 and 2.4 in (\ref{DEC}),
assuming the inequality in Lemma 2.4, we have
\begin{eqnarray}\lefteqn{
\tilde{H}^{(C,Ex)}  \prod_{\gamma=1}^p \Delta (z^{(\gamma)}) \, m_\kappa (w) = }
 \nonumber \\ & & \bigg ( \delta^{(C)} +
{N - 1 -2N^{(z)} + p \over \alpha}  \sum_{j=1}^{N_0}
 w_j {\partial \over \partial w_j}  + \Big ( 1 + {p \over \alpha} \Big )
\bigg [ \sum_{j=1}^{N_0} \Big ( w_j {\partial \over \partial
w_j} \Big )^2 \nonumber \\
& &  + {2 \over \alpha + p} \sum_{j \ne k} {w_jw_k \over w_j - w_k}
{\partial \over \partial w_j} \bigg ]\bigg ) \,  \Delta (z^{(\gamma)}) \,
m_\kappa (w)
\label{WW}
\end{eqnarray}
The first two terms on the r.h.s.~of (\ref{WW}) are eigenoperators of any
homogeneous polynomial in $w$, while the terms in the square brackets
form the eigenoperator defining the symmetric Jack polynomial
$J_\rho^{(p+ \alpha)}(w)$ (recall (\ref{HCC}) with $M_{jk}=1$). Thus by forming
an appropriate linear combination of $m_\kappa(w)$ in (\ref{WW})
(with $|\kappa| = |\rho|$ and ${\rm min}(N_1,\dots,N_p) \ge \rho_1 \ge
\kappa_1$) we see that indeed (\ref{CON}) is an eigenfunction of (\ref{HCC}),
as required.

We remark that the above derivation shows that if we replace 
$J_\rho^{(p+ \alpha)}(w)$ in (\ref{CON}) by $E_\rho(w,\alpha+p)$, then the
resulting function is also an eigenfunction of (\ref{HCC}). This is consistent
with the construction (\ref{DEFS}) of $S_{(\rho,\mu)}$. In fact this latter
eigenfunction must result from antisymmetrizing $E_{(\rho,\mu)}$, with 
$(\rho,\mu)$ defined by (\ref{PK}), in the variables $\{z_j^{(\gamma)}\}_{
j=1,\dots,N_\gamma}$. Thus, with this operation defined by ${\cal A}$ and
 assuming the inequality (\ref{IN}), we see from the structure (\ref{nj})
and the fact that antisymmetrization of a monomial  with equal exponents
vanishes that
\begin{eqnarray}
{\cal A} E_{(\rho,\mu)} & = & w^\rho {\cal A} z^{\mu} +
\sum_{\nu < \rho} c_{\nu \rho} w^{\nu} {\cal A} z^{\mu} \nonumber \\
& = & \prod_{\gamma = 1}^p \Delta(z^{(\gamma)}) \Big (
w^\rho + \sum_{\nu < \rho} c_{\nu \rho} w^{\nu} \Big ).
\label{UNI}
\end{eqnarray}
for some constants $c_{\nu \rho}$.
But ${\cal A} E_{(\rho,\mu)}$ must be an eigenfunction of (\ref{HCC}), and
the above working gives that the function of $w$ must satisfy a eigenvalue
equation in which the eigenoperator is again of the form (\ref{HCC}),
which we know has a unique solution of the form required in (\ref{UNI}).

In the case $q=0$, $p=1$ we can also provide a formula for $S_{(\rho,\mu)} =:
S_\mu$ in terms of the symmetric Jack polynomial:
\begin{equation}
S_\mu (z,\alpha) = \Delta(z) {1 \over v_{\kappa \kappa}(\alpha/(1+\alpha))}
J_\kappa^{(\alpha/(1+\alpha))} (z)
\label{ANTIS}
\end{equation}
where
\begin{equation}
\kappa := (\mu_1 - N+1, \mu_2 - N+2, \dots, \mu_N)
\end{equation}
and $v_{\kappa \kappa}$ is as in (\ref{VK}). Note from
(\ref{DEFS}) that (\ref{ANTIS}) is equivalent to the statement that
\begin{equation}
{\cal A} \, E_\mu(z,\alpha) = \Delta(z) {1 \over v_{\kappa
\kappa}(\alpha/(1+\alpha))} J_\kappa^{(\alpha/(1+\alpha))} (z),
\end{equation}
where ${\cal A}$ denotes antisymmetrization in all variables. The easiest way to
verify (\ref{ANTIS}) is to try for eigenfunctions of (\ref{HCE}) of the form
$|\Delta(z)|^{1/\alpha} \Delta(z) f$ where $f$ is symmetric. A straightforward
calculation  shows that $f$ must be an eigenfunction of (\ref{HCC}) with
$M_{jk}=1$ and $2/\alpha$ replaced by $2/\alpha + 1$. But the unique
symmetric eigenfunction of this equation with leading term $m_\kappa$ is
$ J_\kappa^{(\alpha/(1+\alpha))}/v_{\kappa \kappa}(\alpha/(1+\alpha))$. 
The relationship between $\kappa$ and $\mu$,
and thus the result follows from (\ref{DEFS}).

\section{Eigenfunctions of $\tilde{H}^{(H,Ex)}$}
\setcounter{equation}{0}
\renewcommand{\theequation}{\thesection.\arabic{equation}}
\subsection{The non-symmetric generalized Hermite polynomials}
The operator (\ref{HTHE}) has unique polynomial eigenfunctions of the form
\begin{equation}
y^\eta + \sum_{|\nu| < |\eta|} c_{\eta \, \nu} y^\nu
\label{3.1}
\end{equation}
with corresponding eigenvalue $-2|\eta|$. By adding together an appropriate
linear combination of these eigenfunctions, we can construct the unique
eigenfunction of the form
\begin{equation}
E_\eta^{(H)}(y,\alpha) := E_\eta(y,\alpha) +  \sum_{|\nu| < |\eta|} c_{\eta \,
\nu}' E_{\nu}(y;\alpha)
\label{3.2}
\end{equation}
again with eigenvalue $-2|\eta|$. We will refer to the $E_\eta^{(H)}(y,\alpha)$
as the non-symmetric generalized Hermite polynomials (they are related to the
symmetric generalized Hermite polynomials defined in \cite{forr96a} by an
equation analogous to (\ref{JNSJ}); see  eq.~(\ref{A2}) below). In fact by
adopting a method due to Sogo \cite{sogo96}, an exponential 
operator formula can be obtained expressing $E_\eta^{(H)}(y,\alpha)$ 
in terms of $ E_\eta(y,\alpha)$,
which is the analogue of the formula due to Lassalle \cite{lass96a} (see
eq.~(\ref{A2}) below)
expressing the symmetric generalized Hermite polynomials in terms of the
symmetric Jack polynomials.

To obtain this formula, we write the eigenvalue equation for the
$E_\eta^{(H)}(y,\alpha)$ in the form
\begin{equation}
( A + \tilde{D}_0) E_\eta^{(H)}(y,\alpha) = 0
\label{D0}
\end{equation}
where
\begin{equation}
A := -2 \sum_{j=1}^N y_j {\partial \over \partial y_j} + 2 |\eta|,
\quad
 \tilde{D}_0 :=   \sum_{j=1}^N
 {\partial^2 \over \partial y_j^2 }  + {2 \over \alpha} \sum_{j < k}
{1 \over y_j - y_k} \left[ \Big ( {\partial \over \partial y_j } -
 {\partial \over \partial y_k } \Big ) - {1 - M_{jk} \over y_j - y_k}
\right]
\end{equation}
Note that (\ref{D0}) only specifies $ E_\eta^{(H)}$ uniquely after the
specification (\ref{3.2}).  Since $ E_\eta(y,\alpha)$ is homogeneous of degree
$|\eta|$ we also have
$A \,  E_\eta(y,\alpha) = 0$ which can be equated with (\ref{D0}) and the
resulting equation rearranged to give
\begin{equation}
E_\eta^{(H)}(y,\alpha) = \Big (1 - (A +  \tilde{D}_0)^{-1}\tilde{D}_0 \Big ) \, 
 E_\eta(y,\alpha)
\label{EX1}
\end{equation}

Next we make use of the operator identity
\begin{equation}
 (A +  \tilde{D}_0)^{-1}\tilde{D}_0 =
 A^{-1} \tilde{D}_0 - (A^{-1}\tilde{D}_0)^2 + (A^{-1}\tilde{D}_0)^3 + \dots,
\label{OID}
\end{equation}
We note that after $p$ applications of $\tilde{D}_0$, $ E_\eta(y,\alpha)$ is
a homogeneous polynomial of degree $|\eta| - 2p$ so we have
$$
A^{-1} (\tilde{D}_0)^p \,  E_\eta(y,\alpha) = {1 \over 4 p}  (\tilde{D}_0)^p \, 
E_\eta(y,\alpha).
$$
Using this in (\ref{EX1}) gives
\begin{eqnarray}
E_\eta^{(H)}(y,\alpha) & = & \left( 1 - {1 \over 4}\tilde{D}_0 +
{1 \over 4^2} {1 \over 2!}(\tilde{D}_0)^2 - {1 \over 4^3} {1 \over 3!}
(\tilde{D}_0)^3 + \dots \right)  E_\eta(y,\alpha) \nonumber \\
& = & \exp ( - \tilde{D}_0 / 4 ) \,  E_\eta(y,\alpha),
\label{EX2}
\end{eqnarray}  
which is consistent with (\ref{3.2}) and is thus the sought exponential operator
formula. Note that the series in (\ref{EX2}) terminates after the
$[|\eta|/2]$ application of $\tilde{D}_0$.

To proceed further we note that the operator $ \tilde{D}_0$ can be
written in terms of the Dunkl operator (\ref{DU}) with the $z_i$
replaced by $y_i$. We have
\cite{dunkl89a}
\begin{equation}
 \tilde{D}_0 = \sum_{j=1}^N T_j^2.
\label{DU1}
\end{equation}
We will use (\ref{EX2}) and (\ref{DU1}) to verify that the $E_\eta^{(H)}$ are
simultaneous eigenfunctions of a set of operators more basic than
$\tilde{H}^{(H,Ex)}$, which play an analogous role to the Cherednik operators
in the theory of the non-symmetric Jack polynomials.

\vspace{.2cm}
\noindent
{\bf Proposition 3.1} \quad The non-symmetric generalized Hermite polynomials
$E_\eta^{(H)}$ are eigenfunctions of the operators
\begin{equation}
h_i := \xi_i - {\alpha \over 2} T_i^2, \qquad (i=1,\dots,N)
\label{HI}
\end{equation}
with corresponding eigenvalue $\bar{\eta}_i$. Here $ \xi_i$ is the
Cherednik operator (\ref{COV})
and $\bar{\eta}_i$, which is given explicitly by (\ref{EVCV}), is defined as the
eigenvalue in the  eigenvalue equation
\begin{equation}
\xi_i E_\eta = \bar{\eta}_i  E_\eta.
\label{XEV}
\end{equation}

\vspace{.2cm}
\noindent
{\bf Remarks} \\
(i)  In \cite[Prop.~3.2 with $j=1$]{forr96a} we noted that $D_N^1 + {1 \over 4}
[D_N^1,D_0]$, where
\begin{equation}
D_N^1 := \alpha \sum_{j=1}^N y_j {\partial \over \partial y_j} +N(N-1)/2,
\qquad D_0 :=  \sum_{j=1}^N  {\partial^2 \over \partial y_j^2}
+ {2 \over \alpha} \sum_{j \ne k} {1 \over y_j - y_k} {\partial \over \partial
y_j},
\label{DEFD0}
\end{equation}
is an eigenoperator for the symmetric generalized Hermite polynomials
$H_\kappa(y;\alpha)$. Our construction of (\ref{HI}) was motivated by
 this result, (\ref{DU1})
and the fact that
\begin{equation}
\sum_{i=1}^N (\xi_i + (N-1)) = D_N^1.
\label{SD}
\end{equation}
(ii) We could replace $\xi_i$ in (\ref{HI}) by $\hat{D}_{N+1-i}$ (recall
the remark below (\ref{COV})). Our use of $\xi_i$ has been influenced by
\cite{knop96c,sahi96a}.

\vspace{.2cm}
In further preparation for proving Proposition 3.1 we will evaluate the
commutator $[\xi_j,\tilde{D}_0]$. Due to (\ref{DU1}), we should 
first consider the commutator $[\xi_j,T_i]$.

\vspace{.2cm}
\noindent
{\bf Lemma 3.1} \quad We have
\begin{eqnarray*}
{[}\xi_j, T_i] & = & T_i M_{ij}, \qquad i < j \\
{[}\xi_j, T_i] & = & T_j M_{ij}, \qquad i > j \\
{[}\xi_j, T_j] & = & -\alpha T_j - \sum_{p < j} M_{jp} T_j - \sum_{p> j} T_j
M_{jp}.
\end{eqnarray*}

\vspace{.2cm}
\noindent
{\bf Proof} \quad These formulas are verified by straightforward calculation
using the formula (\ref{COV}) relating $\xi_j$ and $T_j$, the commutator
formula (\ref{DOC}), and the additional
easily verified commutator identities
$$
{[}T_i,y_i]  = 1 + {1 \over \alpha} \sum_{p \ne i} M_{ip} \qquad
{[}T_i,y_j]  =  - {1 \over \alpha} M_{ij}, \quad i \ne j \qquad
{[T}_i, M_{jk}]  =  0, \quad i \ne j,k.
$$

\vspace{.2cm}
Now we can evaluate the commutator  $[\xi_i,\tilde{D}_0]$.

\vspace{.2cm}
\noindent
{\bf Lemma 3.2} \quad We have
$$
[\xi_i,\tilde{D}_0] = - 2 \alpha T_i^2.
$$

\vspace{.2cm}
\noindent
{\bf Proof} \quad Using (\ref{DU1}) we have
$$
[\xi_i,\tilde{D}_0] = \sum_{j=1}^N [\xi_i, T_j^2] =
\sum_{j=1}^N \Big ( [\xi_i, T_j] T_j + T_j[\xi_i, T_j] \Big ).
$$
The result follows after splitting the sum up into parts $j<i$, $j=i$ and
$j > i$, then using Lemma 3.1 and the facts that
$$
M_{ij} T_j = T_i M_{ij}, \qquad M_{ij}T_k = T_k M_{ij} \quad (k \ne i,j).
$$

\vspace{.2cm}
With this preparation we can now provide the verification of the claim
of Proposition 3.1.

\vspace{.2cm}
\noindent 
{\bf Proof of Proposition 3.1} \quad Using (\ref{EX2}) and (\ref{XEV}) we
have
\begin{equation}
\xi_i \Big ( e^{\tilde{D}_0/4} E_\eta^{(H)} \Big ) = \bar{\eta}_i
 \Big ( e^{\tilde{D}_0/4} E_\eta^{(H)} \Big ).
\label{XHEV}
\end{equation}
But according to the Baker-Campbell-Hausdorff formula
\begin{eqnarray}\xi_i \left( e^{\tilde{D}_0/4} E_\eta^{(H)} \right)
& = & e^{\tilde{D}_0/4} \left( \xi_i + {1 \over 4}[\xi_i,\tilde{D}_0]
+{1 \over 2!}{1 \over 4^2} [[\xi_i,\tilde{D}_0],,\tilde{D}_0] + \dots \right) 
E_\eta^{(H)}
\nonumber \\
& = &  e^{\tilde{D}_0/4}  \left( \xi_i -{\alpha \over 2} T_i^2
\right)  E_\eta^{(H)}
\label{BCH}
\end{eqnarray}
where to obtain the last line we have used the fact that since
$[\xi_i,\tilde{D}_0] = -2
\alpha T_i^2$ (by Lemma 3.2) and $\tilde{D}_0 = \sum_{j=1}^N T_j^2$
(eq.~(\ref{DU1})), the higher order commutators vanish due to (\ref{DOC}).
Equating the r.h.s.~of (\ref{BCH}) with the r.h.s.~of (\ref{XHEV}) gives the
desired eigenvalue equation.

\vspace{.2cm}
\noindent
{\bf Remark} \quad Since $\{E_\eta^{(H)}\}$ form a basis for analytic functions
it follows from Proposition 3.1 that $\{h_i\}$ mutually commute. This fact
can also be checked directly using, Proposition 3.1, Lemma 3.1 and (\ref{DOC1})
and (\ref{DOC}).

\vspace{.2cm}
{}From Proposition 3.1, (\ref{DU1}), (\ref{SD}) and (\ref{HTHE}) we have that
\begin{equation}
\sum_{i=1}^N h_i = -{\alpha \over 2} \Big ( \tilde{H}^{(H,Ex)} + N(N-1)/\alpha
\Big ).
\label{A1}
\end{equation}
Also, by forming the sum 
(\ref{JNSJ}) in (\ref{EX2}) we have
\begin{equation}
\sum_{P} a_{P^{-1}\kappa} E_{P^{-1}\kappa}^{(H)}(y;\alpha)  =   \exp \Big ( -{1
\over 4}
\tilde{D}_0
\Big )J_\kappa^{(\alpha)}(y) = 2^{-|\kappa|}J_\kappa^{(\alpha)}(1^N)
H_\kappa(y;\alpha) 
\label{A2}
\end{equation}
where, after noting that $\tilde{D}_0 = D_0$ as defined in (\ref{DEFD0})
when acting on symmetric functions, the last
equality is the exponential operator formula of Lassalle
(see ref.~\cite[eq.~(3.21)]{forr96a}). Now from the remark below Lemma 2.1 
we have that 
$ \{ \bar{\eta}_j \}_{j=1,\dots,N} = \{ \alpha \kappa_j - (j-1)
\}_{j=1,\dots,N}$ independent of the permutation relating $\eta$ to the
partition $\kappa$. Using this fact, (\ref{A1}), (\ref{A2}) and Proposition 3.1
we see by following the argument of the last two paragraphs of Section 2.1
that
\begin{equation}
\prod_{j=1}^N ( 1+ u h_i )
\label{KAKEI}
\end{equation}
is an eigenoperator of the symmetric generalized Hermite polynomials
$H_\kappa(y;\alpha)$ (an operator with this property equivalent to 
(\ref{KAKEI}) has recently been identified by Kakei \cite{kakei96}) 
with corresponding eigenvalue
\begin{equation}
\prod_{j=1}^N \Big ( 1 + u(\alpha \kappa_j - (j-1)) \Big ).
\label{EVH}
\end{equation}
Note that the inequalities (\ref{OR}) imply that the eigenvalues are distinct.

We remark that in ref.~\cite[Prop.~3.2]{forr96a} an 
operator $\tilde{H}_j^{(H)}$ was constructed such that
\begin{equation}
\Big ( \sum_{j=0}^N X^{N-j} \tilde{H}_j^{(H)} \Big ) H_\kappa(y;\alpha)
= \prod_{i=1}^N ( X+ N - i + \alpha \kappa_i)  H_\kappa(y;\alpha).
\label{NH1}
\end{equation}
Comparison with the eigenvalue (\ref{EVH}) corresponding to the operator
(\ref{KAKEI}) shows that when acting on symmetric functions
\begin{equation}
\prod_{i=1}^N ( X+ N + h_i) =  \sum_{j=0}^N X^{N-j} \tilde{H}_j^{(H)}
\label{NH2}
\end{equation}

The eigenoperator (\ref{KAKEI}) can be used to establish that
$\{ H_\kappa(y;\alpha) \}$ are orthogonal with respect to the inner product
\begin{equation}
\langle f|g \rangle^{(H)} := \prod_{l=1}^N \int_{-\infty}^\infty
dy_l \, e^{-y_l^2} \prod_{1 \le j < k \le N} |y_k - y_j|^{2/\alpha} f g
\label{INH}
\end{equation}
(note that the weight function in (\ref{INH}) is equal to $e^{-\beta W^{(H)}}$,
where $ W^{(H)}$ is given by (\ref{WH}), and is thus the square of the
ground state wave function of (\ref{HHE})). This is an immediate consequence
of the fact that the eigenvalues (\ref{EVH}) are distinct and (\ref{KAKEI})
is self-adjoint with respect to (\ref{INH}) (recall the analogous argument in
Section 2.1). The latter result follows from the $h_i$ being self-adjoint with
respect to (\ref{INH}), which is to be established in the subsequent lemma.
The orthogonality has previously been established in 
refs.~\cite{forr96a,kakei96}, but the details here are different.

\vspace{.2cm}
\noindent
{\bf Lemma 3.3} \quad We have
\begin{equation}
\langle f|T_ig \rangle^{(H)} = \langle(2y_i - T_i) f|g \rangle^{(H)}
\label{DU4}
\end{equation}
and thus
\begin{equation}
\langle f|h_ig \rangle^{(H)} = \langle h_i f|g \rangle^{(H)}
\label{DU5}
\end{equation}

\vspace{.2cm}
\noindent
{\bf Proof} \quad The result (\ref{DU4}) is given in 
\cite[lemma 3.7]{dunkl91a}.
It is derived using integration by parts. Using (\ref{DU4}) and the first
equation in (\ref{COV}) we find
\begin{equation}
\langle f|\xi_ig \rangle^{(H)} = 
\langle (\xi_i + \alpha(2y_i^2 - T_i y_j - y_j T_i) f|g \rangle^{(H)}.
\label{DU6}
\end{equation}
But from  Proposition 3.1 $h_i = \xi_i - {(\alpha /2)}T_i^2$.
Noting from (\ref{DU4}) that
\begin{equation}
\langle f|T_i^2g \rangle^{(H)} = \langle(2y_i - T_i)^2 f|g \rangle^{(H)}
\label{DU7}
\end{equation}
the result (\ref{DU5}) follows by subtracting $\alpha/2$ times (\ref{DU7})
from (\ref{DU6}).

Analogous to the theory of the symmetric Jack polynomials revised in Section
2.1, the orthogonality of the symmetric generalized Hermite polynomials can be
established from the formula (\ref{A2}) and the fact that the non-symmetric
generalized Hermite polynomials are orthogonal with respect to (\ref{INH}).
This latter fact is established by noting that (\ref{PONS}) with
$\hat{D}_l$ replaced by $h_l$ is self-adjoint with respect to (\ref{INH})
and is an eigenoperator of each $E_\eta^{(H)}$ which separates the eigenvalues.

\subsection{Generalized Hermite polynomials with prescribed symmetry}
Let $S_{(\rho, \mu)}(y,\alpha)$ denote a Jack polynomial with prescribed
symmetry. By following the working which led to (\ref{EX2}) we can construct a
polynomial eigenfunction of (\ref{HTHE}) according to
\begin{equation}
S_{(\rho, \mu)}^{(H)}(y,\alpha) = \exp( - \tilde{D}_0/4) \,S_{(\rho,
\mu)}(y,\alpha).
\label{EXH}
\end{equation} 
 Note that $S_{(\rho, \mu)}^{(H)}(y,\alpha)$ has
the same symmetry properties as $S_{(\rho, \mu)}(y,\alpha)$. We will refer to
$\{ S_{(\rho, \mu)}^{(H)}(y,\alpha) \}$ as the generalized Hermite polynomials
with prescribed symmetry. Due to the expansion (\ref{SSUM}), and the formula
(\ref{EX2}) we see from (\ref{EXH}) that
\begin{equation}
S_{(\rho, \mu)}^{(H)}(y,\alpha) =
\sum_{\rm rearrangements} b_{Q^{-1}\kappa}
E_{Q^{-1}\kappa}^{(H)}(y,\alpha)
\label{NH3}
\end{equation}

{}From this formula we can deduce that the operator
(\ref{PCO}) with each operator $\hat{D}_j$ replaced by $h_j$ is an
eigenoperator of $S_{(\rho, \mu)}^{(H)}$, and the corresponding eigenvalues are
distinct for distinct members of $\{S_{(\rho, \mu)}^{(H)} \}$. This implies 
$\{S_{(\rho, \mu)}^{(H)} \}$ is an orthogonal set with respect to the inner
product (\ref{INH}). This fact can also be deduced from (\ref{NH3}) and the
orthogonality of $\{ E_\eta^{(H)} \}$.  

\subsection{Some special generalized Hermite polynomials with prescribed
symmetry}
The analogue of the conjecture (\ref{CON}) in the Hermite case is that for
$\eta =: P\kappa $ given by (\ref{PK})
\begin{equation}
S_\eta^{(H)}(y,\alpha) =  A_{\eta}^{(H)}
\prod_{\gamma=1}^p \prod_{1 \le j < k \le N_\gamma}\! (y_k^{(\gamma)} -
y_j^{(\gamma)}) H_{\rho}(\sqrt{\alpha \over \alpha +
p}x_1,\dots,\sqrt{\alpha \over \alpha +p}x_{N_0};p +
\alpha),
\label{CON1}
\end{equation}
provided the inequality (\ref{IN}) is satisfied. To verify this conjecture, our
task is to show that (\ref{CON1}) is an eigenfunction of (\ref{HTHE}). To do
this, we proceed as in (\ref{DEC}) and write  (\ref{HTHE}) when acting on
functions symmetric in $\{x_j\}_{j=1,\dots,N_0}$ and antisymmetric in
$\{y_j^{(\gamma)}\}_{j=1,\dots,N_\gamma}$ $(\gamma = 1,\dots, p)$ as the sum of
three terms:
\begin{equation}
\tilde{H}^{(H,Ex)} = \tilde{H}^{(H,x)} + \tilde{H}^{(H,y)} + \tilde{H}^{(H,xy)}
\end{equation}
where
\begin{equation}
 \tilde{H}^{(H,x)} = \sum_{j=1}^{N_0} \Big ( {\partial^2 \over \partial x_j^2}
 - 2 x_j {\partial \over \partial x_j} \Big ) + {2 \over \alpha}
\sum_{j \ne k} {1 \over x_j - x_k} {\partial \over \partial x_k}
\end{equation}
\begin{equation}
 \tilde{H}^{(H,xy)} = {2 \over \alpha} \sum_{\gamma = 1}^p  \sum_{j=1}^{N_\gamma}
\sum_{k=1}^{N_0} {1 \over y_j^{(\gamma)} - x_k}
 \left[\Big ({\partial \over \partial y_j^{(\gamma)}} -{\partial
\over 
\partial x_k}\Big ) - {1 - M(y_j^{(\gamma)},x_k) \over y_j^{(\gamma)} -
x_k} \right]
\end{equation}
and $\tilde{H}^{(H,y)}$ is given by (\ref{HTHE}) with $y_1,y_2\dots,y_N$
replaced by $y_1^{(1)},y_2^{(1)},\dots,y_{N_p}^{(p)}$.

The fundamental operator in establishing that (\ref{CON1}) is an eigenfunction
of (\ref{HTHE}) is $(y_jy_k)^{-1}A(y_j,y_k)$, where $A(y_j,y_k)$, along with
its action on $y_j^{\lambda_j} y_k^{\lambda_k}$ is specified in Lemma 2.2.
Using this action, we can repeat the argument of the proof of Lemma 2.3
to conclude that $\prod_{\gamma = 1}^p \Delta(y^{(\gamma)})$ is an
eigenfunction of $\tilde{H}^{(H,y)}$. This action can also be used to establish
the analogue of Lemma 2.4.

\vspace{.2cm}
\noindent
{\bf Lemma 3.4}
\quad Let $F(x,y) = m_\kappa (x)  \prod_{\gamma=1}^p \Delta
(y^{(\gamma)})$, where $\kappa$ is a partition consisting of $N_0$ parts
with the largest part $\kappa_1$ restricted by $\kappa_1 \le
{\rm min}(N_1,\dots,N_\gamma)$.
We have 
\begin{equation}
\tilde{H}^{(H,xy)}F(x,y) = {p \over \alpha} \sum_{j=1}^{N_0} 
{\partial^2
\over
\partial x_j^2} F(x,y).
\label{L34}
\end{equation}

\vspace{.2cm}
\noindent
{\bf Proof} \quad Proceeding as in the proof of Lemma 2.4 we see that for $0
\le \lambda \le N_\gamma$ only the $l=1$ term on the r.h.s.~of (\ref{AA})
contributes to the action of $(y^{(\gamma)}_j x_k)^{-1} A(y^{(\gamma)}_j,x_k)
$ on $(y^{(\gamma)}_j)^{\lambda'} x_k^\lambda$, and this requires $\lambda -
\lambda' \ge 2$. Thus
$$
\sum_{j=1}^{N_\gamma} (y^{(\gamma)}_j x_k)^{-1} A(y^{(\gamma)}_j,x_k)
\Delta
(y^{(\gamma)}) x_k^\lambda = \left( \sum_{j=0}^{\lambda - 2}
(\lambda - j - 1) \right) \Delta
(y^{(\gamma)}) x_k^\lambda,
$$
which implies
$$
\tilde{H}^{(H,xy)}F(x,y) = {p \over \alpha} (|\kappa^2| - |\kappa|) F(x,y),
$$
and the result follows.

\vspace{.2cm}
{}From Lemma 3.4 and the fact that $\tilde{H}^{(H,y)}$ is an eigenoperator for
$\prod_{\gamma = 1}^p \Delta
(y^{(\gamma)})$ with eigenvalue $\delta^{(H)}$ say we have, assuming the
inequality in Lemma 3.4,
\begin{eqnarray*} \tilde{H}^{(H,Ex)} \prod_{\gamma = 1}^p \Delta
(y^{(\gamma)}) m_\kappa(x) & = & \left(\delta^{(H)} + (1 + {p \over
\alpha})
\left[ \sum_{j=1}^{N_0} {\partial^2 \over \partial x_j^2} - {2 \alpha \over
\alpha + p} x_j {\partial \over \partial x_j}\right.\right.\\
&& \left.\left.+ {2 \over \alpha + p}
\sum_{j \ne k}{1 \over x_j - x_k}{\partial \over \partial x_j}
\right] \right) \prod_{\gamma = 1}^p \Delta
(y^{(\gamma)}) m_\kappa(x).
\end{eqnarray*}
The fact that (\ref{CON1}) is an eigenfunction of $ \tilde{H}^{(H,Ex)}$ follows
from this equation since the operator in square brackets is the defining
eigenoperator for $H_\kappa((\alpha/(\alpha + p))^{1/2}x;\alpha)$.

As another explicit evaluation of a class of $S_{(\rho, \mu)}^{(H)}$, in the
case $p=1$, $q=0$ we have
 the analogue of (\ref{ANTIS}):
\begin{equation}
S_\mu^{(H)}(y,\alpha) = \Delta (y) {2^{-|\kappa|} C_\kappa^{(\alpha)}(1^N)
\over v_{\kappa \kappa}(\alpha/(1+\alpha))}
H_\kappa(y;\alpha/(1+\alpha)),
\label{ANTIS1}
\end{equation}
where the prefactors of $H_\kappa$ are chosen so that the coefficient of
$m_\kappa$ in this factor is unity (see ref.~\cite{forr96a}). The derivation of
this result is analogous to the derivation of  (\ref{ANTIS}), and so will be
omitted.

\section{Eigenfunctions of $\tilde{H}^{(L,Ex)}$}
\setcounter{equation}{0}
\renewcommand{\theequation}{\thesection.\arabic{equation}}
\subsection{The non-symmetric generalized Laguerre polynomials}
Analogous to the situation with $\tilde{H}^{(H,Ex)}$, the operator
(\ref{HTLE}) has unique polynomial eigenfunctions of the form (\ref{3.1})
with corresponding eigenvalue $-|\eta|$. An appropriate linear combination of
these eigenfunctions gives unique eigenfunctions of the form
\begin{equation}
E_\eta^{(L)}(y,\alpha) := E_\eta(y,\alpha) +  \sum_{|\nu| < |\eta|} c_{\eta \,
\nu} E_{\nu}(y;\alpha)
\label{4.1}
\end{equation}
which also have eigenvalue $-|\eta|$. 
As well as depending on the parameter $\alpha$ they also depend on the
parameter $a$ in (\ref{HTLE}), however for notational convenience we have
suppressed this dependence in (\ref{4.1}). The
$E_\eta^{(L)}(y,\alpha)$ will be referred to as the non-symmetric generalized
Laguerre polynomials (we recall from ref.~\cite{forr96a} that the symmetric
generalized Laguerre polynomials
$L_\kappa^a(y;\alpha)$ are the polynomial eigenfunctions of 
(\ref{HTLE}) with leading term proportional to the symmetric Jack
polynomial $J_\kappa^{(\alpha)}(y)$.

By repeating the working which led to (\ref{EX2}), starting with the eigenvalue
equation for $E_\eta^{(L)}(y,\alpha)$, we obtain the exponential operator
formula
\begin{equation}
E_\eta^{(L)}(y,\alpha) = \exp \bigg ( - \Big ( \tilde{D}_1 +
(a+1) \sum_{j=1}^N {\partial \over \partial y_j} \Big ) \bigg ) E_\eta(y,\alpha)
\label{EX5}
\end{equation}
where
\begin{equation}
\tilde{D}_1 := \sum_{j=1}^N y_j {\partial^2 \over \partial y_j^2} +
{1 \over \alpha} \sum_{j<k} {1 \over y_j - y_k} \left[ 2\Big (
y_j {\partial \over \partial y_j} - y_k {\partial \over \partial y_k}\Big ) -
{y_j + y_k \over y_j - y_k}(1 - M_{jk}) \right].
\label{WE1}
\end{equation}

It is convenient to introduce new variables $x_j^2 =: y_j$. It follows from
\cite[first eq.~pg.~125]{dunkl93a} that
\begin{equation}
 \tilde{D}_1 +
\left.(a+1) \sum_{j=1}^N {\partial \over \partial y_j} \right|_{y_j \mapsto
x_j^2} = {1 \over 4} \sum_{i=1}^N \left(T_i^{(B)}\right)^2
\label{WE2}
\end{equation}
where $T_i^{(B)}$ is the Dunkl operator for the root system $B_N$:
\begin{equation}
T_i^{(B)} := { \partial \over \partial x_i} + {1 \over \alpha} \sum_{p \ne i}
\left( { 1 - M_{ip} \over x_i - x_p} + {1 - S_i S_p M_{ip} \over x_i + x_p}
\right) + {a + 1/2 \over x_i}(1 - S_i)
\label{DEFTB}
\end{equation}
(as in (\ref{HLE}) $S_j$ is the operator which replaces the coordinate $x_j$ by
$-x_j$). The similarity between (\ref{EX5}) with the substitution (\ref{WE2}),
and (\ref{EX2}) with the substitution (\ref{DU1}) suggest we define  operators
$l_i$ say, analogous to (\ref{HI}):
\begin{equation}
l_i := \hat{\xi}_i - {\alpha \over 4} (T_i^{(B)})^2
\label{LI}
\end{equation}
where $ \hat{\xi}_i$ is the Cherednik operator (\ref{COV}) with the change of
variables $y_j = x_j^2$ (a literal analogy would have $\alpha/4$ replaced by
$\alpha/2$ in (\ref{LI}); the reason for modifying this is connected with the
remark accompanying Lemma 4.2 below). We want to show that the
$l_i$ are eigenoperators for the $E_\eta^{(L)}(x^2,\alpha)$. To do this we
require the analogue of Lemma 3.2, and this in turn requires some preliminary
results.

\vspace{.2cm}
\noindent
{\bf Lemma 4.1} \quad When acting on functions even in $x_1,\dots,x_N$
\begin{equation}
{1 \over 4}  (T_i^{(B)})^2 = x_i^2 \hat{T}_i^2 + (a+1) \hat{T}_i +
{1 \over \alpha} \sum_{p \ne i} M_{ip} \hat{T}_i
\label{L41}
\end{equation}
where $ \hat{T}_i$ is the $A$-type Dunkl operator (\ref{DU}) with the
change of variables $z_j = x_j^2$ $(j=1,\dots,N)$:
\begin{equation}
\hat{T}_i = {1 \over 2x_i} { \partial \over \partial x_i}
+ {1 \over \alpha} \sum_{p \ne i} {1 - M_{ip} \over x_i^2 - x_p^2}.
\label{DEFTH}
\end{equation}

\vspace{.2cm}
\noindent
{\bf Proof} \quad Let $f$ be even in $x_1,\dots,x_N$. Then from
(\ref{DEFTB}) and (\ref{DEFTH}) we see that
$$
T_i^{(B)} f = 2x_i \hat{T}_i \, f.
$$
Now $x_i \hat{T}_i \, f$ is odd in $x_i$, and from the definition
(\ref{DEFTB}) we see that when acting on a function which is odd in $x_i$,
$$
T_i^{(B)} = 2x_i  \hat{T}_i + {2a+1 \over x_i} + {1 \over \alpha}
\sum_{p \ne i} {1+S_p \over x_i + x_p} M_{ip}.
$$
Thus when acting on $f$
\begin{equation}
 (T_i^{(B)})^2  =  4 (x_i  \hat{T}_i)^2 + (4a+2) \hat{T}_i
+ {4 \over \alpha} \sum_{p \ne i} {1 \over x_i + x_p} M_{ip}x_i  \hat{T}_i.
\label{PL41}
\end{equation}
To simplify further, note that
\begin{equation}
 (x_i  \hat{T}_i)^2 = x_i ( [ \hat{T}_i,x_i] + x_i   \hat{T}_i ) \hat{T}_i
\label{PL42}
\end{equation}
and evaluate the commutator:
\begin{equation}
[ \hat{T}_i,x_i] = {1 \over 2 x_i} + {1 \over \alpha} \sum_{p
\ne i} {1 \over x_i + x_p} M_{ip}.
 \label{PL43}
\end{equation}
The stated result (\ref{L41}) follows by substituting (\ref{PL43}) in
(\ref{PL42}), substituting the result in (\ref{PL41}) and simplifying.

\vspace{.2cm}
The analogue of Lemma 3.1 is given by the following result.

\vspace{.2cm}
\noindent
{\bf Lemma 4.2} \quad Let $B_i := {1 \over 4}(T_i^{(B)})^2$ and let
\begin{equation}
 \hat{\xi}_j := \alpha x_j^2  \hat{T}_j + (1 - N) + \sum_{p > j} M_{jp}
\label{DEFXH}
\end{equation}
be the Cherednik operator (\ref{COV}) with the substitution $y_j = x_j^2$
($j=1,\dots,N$). We have
\begin{eqnarray*}
[ \hat{\xi}_j,B_i] & = & B_i M_{ij}, \quad i<j \\
{[} \hat{\xi}_j,B_i] & = & B_j M_{ij}, \quad i>j \\
{[} \hat{\xi}_j,B_j]& = &- \alpha B_j - \sum_{p < j} M_{jp} B_j - \sum_{p > j}
B_j M_{jp}.
\end{eqnarray*}

\vspace{.2cm}
\noindent
{\bf Remark} \quad Although the commutators in Lemma 3.1 and Lemma 4.2 have the
same structure, note that in Lemma 3.1 the operator $T_i$ occurs while in
Lemma 4.2 it is the operator $(T_i^{(B)})^2$ which occurs. Moreover, it
seems that there is no simple formula for $[ \hat{\xi}_j,T_i^{(B)}]$.

\vspace{.2cm}
\noindent
{\bf Proof} \quad First consider the case $i < j$. From Lemma 4.1 we have
\begin{eqnarray}
[ \hat{\xi}_j,B_i] & = & [ \hat{\xi}_j,  x_i^2 \hat{T}_i^2 + (a+1) \hat{T}_i +
{1 \over \alpha} \sum_{p \ne i} M_{ip} \hat{T}_i]  \nonumber \\
& = & [ \hat{\xi}_j,  x_i^2] \hat{T}_i^2 + x_i^2 \Big (
[ \hat{\xi}_j, \hat{T}_i] \hat{T}_i +  \hat{T}_i [ \hat{\xi}_j, \hat{T}_i]
\Big ) \nonumber \\
& & + (a+1) [ \hat{\xi}_j, \hat{T}_i] + {1 \over \alpha}
[  \hat{\xi}_j, \sum_{p \ne i} M_{ip} ]  \hat{T}_i + {1 \over \alpha}
\sum_{p \ne i} M_{ip} [ \hat{\xi}_j, \hat{T}_i].
\label{MCOM}
\end{eqnarray}
Now, for $i <j$ a direct calculation gives $ [ \hat{\xi}_j,  x_i^2] = -x_j^2
M_{ij}$, while Lemma 3.1 gives $[ \hat{\xi}_j, \hat{T}_i] =  \hat{T}_i M_{ij}$.
To evaluate the second last commutator in (\ref{MCOM}) we substitute
(\ref{DEFXH}), and note that 
$$
[x_j^2 \hat{T}_j, \sum_{p \ne i} M_{ip}] = (x_j^2 \hat{T}_j - x_i^2 \hat{T}_i)
M_{ij}, \qquad [\sum_{q > j} M_{jq}, \sum_{p \ne i} M_{ip}]=0,
$$
where to obtain the latter result the formula $M_{ij} M_{jp} =
M_{jq}M_{qi}= M_{qi}M_{ij}$ has been used. Substituting these results in
(\ref{MCOM}) gives the stated result in the case $i<j$. The cases
$i >j$ and $i=j$ are very similar, although the latter case requires more
manipulation to simplify the final expression.

\vspace{.2cm}
Using Lemma 4.2  the required analogue of Lemma 3.2 follows.

\vspace{.2cm}
\noindent
{\bf Lemma 4.3} \quad We have
$$
[ \hat{\xi}_j,  \sum_{i=1}^N (T_i^{(B)})^2] = -\alpha 
(T_j^{(B)})^2.
$$

\vspace{.2cm}
\noindent We can use Lemma 4.3 in the same way as Lemma 3.2 was used in the
proof of Proposition 3.1 to prove the analogue of Proposition 3.1 in the
Laguerre case.

\vspace{.2cm}
\noindent
{\bf Proposition 4.1} \quad The non-symmetric Laguerre polynomials
$E_\eta^{(L)}(x^2,\alpha)$ are simultaneous eigenfunctions of the operators 
$l_i$ (\ref{LI}) with corresponding eigenvalue $\bar{\eta}_i$.
 
\vspace{.2cm}
\noindent
{\bf Remark} \quad Since $\{E_\eta^{(L)} \}$ form a basis for analytic
functions, it follows from Proposition 4.1 that $\{l_i\}$ mutually commute, a
fact that can also be derived directly by using Lemma 4.2 and the fact that
the $T_i^{(B)}$ commute, as do the operators $ \hat{\xi}_i$.

\vspace{.2cm}
{}From (\ref{SD}) and (\ref{WE2}) we see from (\ref{LI}) and (\ref{HTLE}) that
\begin{equation}
\sum_{i=1}^N l_i =\left. -\alpha \tilde{H}^{(L,Ex)} 
\right|_{y_j \mapsto x_j^2} - N(N-1)/2.
\label{LI2}
\end{equation}
Also, analogous to (\ref{A2}), forming the sum (\ref{JNSJ}) in (\ref{EX5})
we have
\begin{eqnarray}
\sum_{P} a_{P^{-1}\kappa} E_{P^{-1}\kappa}^{(L)}(y;\alpha) & = &
 \exp \bigg ( - \Big ( \tilde{D}_1 +
(a+1) \sum_{j=1}^N {\partial \over \partial y_j} \Big ) \bigg )
J_\kappa^{(\alpha)}(y) \nonumber \\
& = & (-1)^{|\kappa|} |\kappa|! J_\kappa^{(\alpha)}(1^N) L_\kappa^a(y;\alpha)
\label{NL1}
\end{eqnarray}
where the last equality follows from \cite[eq.~(4.39)]{forr96a}. From 
(\ref{LI2}), analogous to (\ref{KAKEI}), we have that
\begin{equation}
\prod_{j=1}^N ( 1+ u l_i )
\label{KAKEI2}
\end{equation}
is an eigenoperator of the symmetric generalized Laguerre polynomials
$ L_\kappa^a(x^2;\alpha)$ with corresponding eigenvalue (\ref{EVH}).

In ref.~\cite[Prop.~4.5]{forr96a} an operator $\tilde{H}_j^{(L)}$ was 
constructed
such that (\ref{NH1}) holds with $\tilde{H}_j^{(H)}$ replaced by
$\tilde{H}_j^{(L)}$ and $H_\kappa(y;\alpha)$ replaced by $L_\kappa^a(y;\alpha)$.
It follows that (\ref{NH2}) holds with $h_j$ replaced by $l_j$ and
$\tilde{H}_j^{(H)}$ replaced by $\tilde{H}_j^{(L)}$.

We know from \cite{forr96a} (see also \cite{lass91b}) that $\{L_\kappa^a\}$
are orthogonal with respect to the inner product
\begin{eqnarray}
\langle f | g \rangle^{(L)} &:=& 2^N \prod_{l=1}^N \int_{-\infty}^\infty dx_l \,
e^{-x_l^2} |x_l|^{2a+1} \prod_{1 \le j < k \le N} |x_k^2 - x_j^2|^{2/\alpha}
\nonumber \\&& \times f(x_1^2,\dots,x_N^2) g(x_1^2,\dots,x_N^2)
\label{INL}
\end{eqnarray}
Note that the weight function is proportional to $e^{-\beta W^{(L)}}$, where
$W^{(L)}$ is given by (\ref{WL}), and is thus proportional to the square of
the symmetric ground state wave function of (\ref{HTLE}).
The orthogonality can be deduced in the present setting by first checking
(see subsequent lemma) that the $l_i$, and thus the eigenoperator
(\ref{KAKEI2}), are self adjoint with respect to (\ref{INL}) and recalling that
the eigenvalues of (\ref{KAKEI2}) are distinct.

\vspace{.2cm}
\noindent
{\bf Lemma 4.4} \quad We have
\begin{equation}
\langle f |\hat{T}_i g \rangle^{(L)} = \langle (1 - {a \over x_i^2} - \hat{T}_i
)f|g 
\rangle^{(L)}
\label{L441}
\end{equation}
and 
\begin{equation}
\langle f | l_i g \rangle^{(L)} = \langle l_i f | g  \rangle^{(L)}.
\label{L442}
\end{equation}

\vspace{.2cm}
\noindent
{\bf Proof} \quad The result (\ref{L441}) follows from the explicit formula
(\ref{DEFTH}) and integration by parts. From (\ref{L441}), (\ref{PL43}) and
(\ref{DEFXH}) we find
\begin{equation} 
\langle f |\hat{\xi}_i g \rangle^{(L)} = \langle (\hat{\xi}_i + \alpha
(x_i^2 - a - 2x_i^2 \hat{T}_i - 1 - {1 \over \alpha} \sum_{p \ne i}
M_{ip}))f|g \rangle^{(L)}.
\label{PL441}
\end{equation}
Also, analogous to (\ref{DU4}) we have
\begin{equation}
\langle f|T_i^{(B)}g \rangle^{(L)} = \langle(2x_i - T_i^{(B)}) f|g
\rangle^{(L)}
\label{PL442}
\end{equation}
and thus
\begin{equation}
\langle f|(T_i^{(B)})^2g \rangle^{(L)} = \langle(4x_i^2 - 2x_iT_i^{(B)}
-2T_i^{(B)} x_i + (T_i^{(B)})^2) f|g
\rangle^{(L)}.
\label{PL443}
\end{equation}
But from the proof of Lemma 4.1 we know that when acting on functions
even in $x_1^2, \dots, x_N^2$,
$$
x_i T_i^{(B)} = 2x_i^2 \hat{T}_i
$$
and from the working in the same proof we can compute that
$$
T_i^{(B)} x_i = 2x_i^2\hat{T}_i + 2(a+1) + {2 \over \alpha}
\sum_{p \ne i} M_{ip}.
$$
Substituting these formulas in (\ref{PL443}) and subtracting $\alpha/4$ times
the result from (\ref{PL441}) gives the required result (\ref{L442}).

Note that the operator (\ref{PONS}) with the $\hat{D}_j$ replaced by $l_j$
is a self-adjoint (with respect to (\ref{INL})) eigenoperator of the
$E_\eta^{(L)}(x^2,\alpha)$ which separates the eigenvalues. It follows that
$\{E_\eta^{(L)} \}$ is an orthogonal set with respect to the inner product
(\ref{INL}). The orthogonality of $\{ L_\kappa^a \}$ also follows from this fact
and the expansion (\ref{NL1}).

\subsection{Generalized Laguerre polynomials with prescribed symmetry}
The theory here is analogous to that for the generalized Hermite
polynomials with prescribed symmetry. Generalized Laguerre polynomials with 
prescribed symmetry,
$S_{(\rho, \mu)}^{(L)}(y,\alpha)$ say, are defined as the eigenfunctions of
(\ref{HTLE}) given by the
 exponential operator formula
\begin{equation}
S_{(\rho, \mu)}^{(L)}(y,\alpha) = \exp \bigg ( - \Big ( \tilde{D}_1 +
(a+1) \sum_{j=1}^N {\partial \over \partial y_j} \Big ) \bigg )
S_{(\rho,\mu)}(y,\alpha).
\end{equation}

The operator (\ref{PCO}) with each $\hat{D}_j$ replaced by $l_j$ is an
eigenoperator for each $S_{(\rho, \mu)}^{(L)}(y,\alpha)$ and the 
eigenvalues are distinct for distinct members of $\{S_{(\rho, \mu)}^{(L)} \}$. 
{}From this we see that 
 $\{S_{(\rho, \mu)}^{(L)} \}$ is an orthogonal set with respect to the inner
product (\ref{INL}).

\subsection{Some special generalized Laguerre polynomials with prescribed
symmetry}
For $\eta =: P^{-1}\kappa$ given by  (\ref{PK}) and the inequality 
 (\ref{IN})  satisfied 
the analogue of the conjecture (\ref{CON}) in the Laguerre case states
\begin{equation}
S_\eta^{(L)}(y,\alpha) =  A_{\eta}^{(L)}
\prod_{\gamma=1}^p \prod_{1 \le j < k \le N_\gamma} (y_k^{(\gamma)} -
y_j^{(\gamma)}) L_{\rho}^{\alpha a /(p+\alpha)}\Big ({\alpha \over \alpha +
p}x_1,\dots,{\alpha \over \alpha +
p}x_{N_0};p +
\alpha \Big ),
\label{CON2}
\end{equation}
Here we will verify this statement by showing that the r.h.s.~of (\ref{CON2})
is an eigenfunction of (\ref{HTLE}). Now, when acting on
functions symmetric in $\{x_j\}_{j=1,\dots,N_0}$ and antisymmetric in
$\{y_j^{(\gamma)}\}_{j=1,\dots,N_\gamma}$ $(\gamma = 1,\dots, p)$ we have
\begin{equation}
\tilde{H}^{(H,Ex)} = \tilde{H}^{(L,x)} + \tilde{H}^{(L,y)} + \tilde{H}^{(L,xy)}
\end{equation}
where
\begin{equation}
 \tilde{H}^{(L,x)} = \sum_{j=1}^{N_0} \Big ( x_j{\partial^2 \over 
 \partial x_j^2}
+(a+1 -  x_j) {\partial \over \partial x_j} \Big ) + {2 \over \alpha}
\sum_{j \ne k} {x_j \over x_j - x_k} { \partial \over \partial x_j} 
\end{equation}
\begin{equation}
 \tilde{H}^{(L,xy)} = {1 \over \alpha} \sum_{\gamma = 1}^p  
 \sum_{j=1}^{N_\gamma}
\sum_{k=1}^{N_0} {1 \over y_j^{(\gamma)} - x_k}
 \left[2\Big ( y_j^{(\gamma)}{\partial \over \partial y_j^{(\gamma)}} -
x_k {\partial
\over 
\partial x_k}\Big ) - {  y_j^{(\gamma)} + x_k \over y_j^{(\gamma)} -
x_k}(1 - M(y_j^{(\gamma)},x_k) \right]
\label{HLXY}
\end{equation}
and $\tilde{H}^{(L,y)}$ is given by (\ref{HTLE}) with $y_1,y_2\dots,y_N$
replaced by $y_1^{(1)},y_2^{(2)},\dots,y_{N_p}^{(p)}$.

To compute the action of these operators we require the action of the operator
which occurs in the summand of (\ref{HLXY}) on monomials. A direct
calculation gives the following result.

\vspace{.2cm}
\noindent
{\bf Lemma 4.4} \quad
Let
$$
A^{(L)}_{jk} := {1 \over y_j - y_k} \left[ 2 \Big (
y_j{\partial \over \partial y_j} - y_k{\partial \over \partial y_k} \Big )
- {y_j + y_k \over y_j - y_k}(1 - M_{jk}) \right].
$$
For $\lambda_j \ge \lambda_k$ we have
$$
A^{(L)}_{jk} y_j^{\lambda_j}  y_k^{\lambda_k} = \left\{
\begin{array}{l} {\displaystyle \sum_{l=1}^{\lambda_j - \lambda_k }}
( 2 (\lambda_j - \lambda_k - l) + 1  ) y_j^{\lambda_j - l}
y_k^{\lambda_k - l + 1}, \quad \lambda_j - \lambda_k \ge 1 \\[2mm]
0, \quad {\rm otherwise}. \end{array} \right.
$$

\vspace{.2cm}
Using Lemma 4.4 the argument of the proof of Lemma 2.3 shows that
$\prod_{\gamma = 1}^p \Delta(y^{(\gamma)})$ is an eigenfunction of
$\tilde{H}^{(L,y)}$.  This lemma is also used to determine the action of
$\tilde{H}^{(L,xy)}$. The strategy is the same as in the proof of Lemma 2.4, so
the details will be omitted.

\vspace{.2cm}
\noindent
{\bf Lemma 4.5} \quad
Let $F(x,y)$ be as in Lemma 3.4. We have
$$
\tilde{H}^{(L,xy)} F(x,y) = {p \over \alpha} \sum_{j=1}^{N_0} \Big (
x_j {\partial^2
\over
\partial x_j^2} + {\partial \over \partial x_j} \Big ) F(x,y).
$$

\vspace{.2cm}
{}From the above working we see that, assuming the inequality in Lemma 3.4,
\begin{eqnarray*}\lefteqn{ \tilde{H}^{(L,Ex)} \prod_{\gamma = 1}^p \Delta
(y^{(\gamma)}) m_\kappa(x)  =  \left( \delta^{(L)} + (1 + {p \over \alpha})
\left[ \sum_{j=1}^{N_0} x_j{\partial^2 \over \partial x_j^2} 
+\Big ( a\alpha/(p+\alpha) +1  \right.\right. }\\
&&\left.\left. -x_j \alpha/(p+\alpha) \Big )
{\partial \over \partial x_j} 
 + {2 \over \alpha + p}
\sum_{j \ne k}{x_j \over x_j - x_k}{\partial \over \partial x_j}
\right] \right) \prod_{\gamma = 1}^p \Delta
(y^{(\gamma)}) m_\kappa(x).
\end{eqnarray*}
where $\delta^{(L)}$ is the eigenvalue for the action of $\tilde{H}^{(L,y)}$.
Since the operator in square brackets is the defining eigenoperator for
$L_\kappa^{\alpha a/(p + \alpha)} (\alpha x/(p + \alpha); \alpha)$, it follows
that (\ref{CON2}) is an eigenfunction of $\tilde{H}^{(L,Ex)}$ as required.

In the
case $p=1$, $q=0$ we have
 the analogue of (\ref{ANTIS}) and (\ref{ANTIS1}):
\begin{equation}
S_\mu^{(L)}(y,\alpha) = \Delta (y) {(-1)^{|\kappa|} |\kappa|!
C_\kappa^{(\alpha)}(1^N)
\over v_{\kappa \kappa}(\alpha/(1+\alpha))}
L_\kappa^a(y;\alpha/(1+\alpha)),
\end{equation}
which is derived according to the same method.

\section{Eigenfunctions of $\tilde{H}^{(J,Ex)}$}
\setcounter{equation}{0}
\renewcommand{\theequation}{\thesection.\arabic{equation}}
\subsection{Decomposition}
The operators, $\hat{D}_j^{(J)}$ say, which provide a decomposition of
$\tilde{H}^{(J,Ex)}$ analogous to the decomposition (\ref{FAC}) of
$\tilde{H}^{(C,Ex)}$ have been given by Bernard et al.~\cite{bern95a} and 
Hikami \cite{hik96a}. We have
\begin{equation}
\hat{D}_j^{(J)} = \hat{D}_j - {1 \over \alpha} \sum_{k=1 \atop \ne j}^N
{1 - S_j S_k M_{jk} \over 1 - z_j z_k} - (a+{1 \over 2})
{1 - S_j \over 1 - z_j} - (b+{1 \over 2}){1 - S_j \over 1 + z_j}
+ {1 \over 2}(a+b+1)
\label{CJ}
\end{equation}
and
\begin{equation}
\tilde{H}^{(J,Ex)} = \sum_{j=1}^N \Big ( \hat{D}_j^{(J)}  \Big )^2
-{1 \over 4} E_0^{(J)}.
\label{HJC}
\end{equation}
We want to investigate the eigenfunctions of $\hat{D}_j^{(J)}$ and relate
them to the eigenfunctions of $\tilde{H}^{(J,Ex)}$. To begin, we note that
$\hat{D}_j^{(J)}$ is self-adjoint with respect to the inner product
\begin{equation}
\langle f | g \rangle^{(J)} := \int_0^{\pi/2} d\phi_1 \dots 
 \int_0^{\pi/2} d\phi_N \, |\psi_0^{(J)}|^2 f(z_1^*,\dots,z_N^*)
g(z_1,\dots,z_N)
\label{IPJ}
\end{equation}
where $z_j = e^{2 i \phi_j}$ and
\begin{equation}
\psi_0^{(J)} := \prod_{j=1}^Nz_j^{-(N-1)/\alpha - (a+b+1)/2}
(z_j-1)^{a+1/2} (z_j + 1)^{b+1/2} \prod_{j<k} (z_j -
z_k)^{1/\alpha} (1 - z_j z_k)^{1/\alpha}.
\end{equation}
Note that $|\psi_0^{(J)}|^2$ is proportional to $e^{-\beta W^{(J)}}$ where
$W^{(J)}$ is given by (\ref{WJ}), and is thus proportional to the square of the
symmetric ground state wave function of $H^{(J,Ex)}$. The self-adjointness is
easily checked upon using the operator identity
\begin{eqnarray}\lefteqn{
\psi_0^{(J)} \hat{D}_j^{(J)} (\psi_0^{(J)})^{-1}  = 
z_j {\partial \over  \partial z_j} -
{1 \over \alpha} \Big ( \sum_{l < j} {z_l \over z_j - z_l}  M_{lj}
+ \sum_{l > j} {z_j \over z_j - z_l}  M_{lj} \Big )}
 \hspace{3cm}\nonumber \\
&&+{1 \over \alpha} \sum_{k=1 \atop \ne j} {S_j S_k M_{jk} \over 1 - z_j
z_k} +   (a+1/2)
{ S_j \over 1 - z_j} + (b+1/2){ S_j \over 1 + z_j}.
\label{REMA}
\end{eqnarray}
Also, we remarked  in a previous study \cite{forr96a}  that
 $\tilde{H}^{(J,Ex)}$ has a complete set of symmetric polynomial
eigenfunctions $G_\kappa^{(a,b)}(y;\alpha)$
(the generalized Jacobi polynomials) where $y=\sin^2 \phi = - (z+1/z-2)/4$.

To specify the eigenvalues and  eigenfunctions of (\ref{CJ}), let $\eta$
be an $N$-tuple of non-negative integers as in  (\ref{ETA}), and define the
partial order as below (\ref{ETA}). Let $\epsilon$ be an $N$-tuple with
each entry $+1$ or $-1$, and define $\epsilon \eta$ as the $N$-tuple
formed from $\epsilon$ and $\eta$ by multiplication of the respective parts.
A direct calculation shows
\begin{equation}
 \hat{D}_j^{(J)} \, z^{\epsilon \eta} 
=e_{j,\epsilon\eta}^{(J)} z^{\epsilon \eta} + \sum_{\eta' < \eta}
\sum_{\epsilon'}  c_{\epsilon \eta, \epsilon' \eta'}z^{\epsilon' \eta'}
\label{CHA} 
\end{equation}
where
\begin{eqnarray}
e_{j,\epsilon\eta}^{(J)} & = & \epsilon_j \eta_j + {1 \over \alpha}
\Big ( - \sum_{l<j} h(\epsilon_l \eta_l - \epsilon_j \eta_j) +
\sum_{l>j} h(\epsilon_j \eta_j - \epsilon_l \eta_l) +
\sum_{k=1 \atop \ne j}^N h(-\epsilon_j \eta_j -\epsilon_k \eta_k) \Big )
\nonumber \\
& & + {1 \over \alpha}(j-1)   - (a+b+1) \Big (
h(-\epsilon_j
\eta_j) - {1 \over 2} \Big )
\label{CEVJ}
\end{eqnarray}
with $h(x)$ defined by (\ref{HX}). It follows that $\hat{D}_j^{(J)}$
 has a complete set of Laurent polynomial eigenfunctions such that
when the highest weight monomial is $z^{\epsilon \eta}$ the corresponding
eigenvalue is $e_{j,\epsilon\eta}^{(J)}$. The fact that each operator
$\hat{D}_j^{(J)} $ has a unique  Laurent polynomial eigenfunction
with leading monomial  $z^{\epsilon \eta}$ and that 
the operators $\{ \hat{D}_j^{(J)} \}$ mutually commute imply each
eigenfunction is simultaneously an eigenfunction of all the operators
$\hat{D}_1^{(J)}, \dots,\hat{D}_N^{(J)}$.

Next we note that the eigenvalues $e_{j,\epsilon\eta}^{(J)}$ and
$e_{j,\epsilon\eta}^{(J)} \Big |_{\epsilon_p \mapsto - \epsilon_p}$ are
simply related.

\vspace{.2cm}
\noindent
{\bf Lemma 5.1} \quad We have
$$
e_{j,\epsilon\eta}^{(J)} \Big |_{\epsilon_p \mapsto - \epsilon_p} =
e_{j,\epsilon\eta}^{(J)} \quad (j \ne p), \qquad
e_{j,\epsilon\eta}^{(J)} \Big |_{\epsilon_p \mapsto - \epsilon_p} =
-e_{j,\epsilon\eta}^{(J)} \quad (j = p).
$$

\vspace{.2cm}
\noindent
{\bf Proof} \quad This follows directly from  (\ref{CEVJ}).

\vspace{.2cm}
\noindent
A consequence of Lemma 5.1 is that $(\hat{D}_j^{(J)})^2$
permits Laurent polynomial eigenfunctions, $E_\eta^{(J)}(y)$ say
($y=-(z+1/z-2)/4$), with leading
term $y^{\eta}$ and corresponding eigenvalue 
$(e_{j,\epsilon\eta}^{(J)})^2 =(e_{j,\eta}^{(J)})^2 $. Now
$e_{j,\eta}^{(J)} = e_{j,\eta} + (a + b +1)/2$ where $e_{j,\eta}$ is
given by (\ref{EV}). From Lemma 2.1 we thus have that
$\{ (e_{j,\eta}^{(J)})^2 \}$ is independent of the permutation in
the equation $\eta = P \kappa$.

By following the argument used in the last two paragraphs of Section 2.1 we
conclude that
\begin{equation}
\Big ( 1 + u (\hat{D}_1^{(J)})^2 \Big )
\dots \Big ( 1 + u (\hat{D}_N^{(J)})^2 \Big )
\label{PEJ}
\end{equation}
is an eigenoperator for the symmetric (Laurent) polynomial eigenfunctions
of (\ref{HJC}) with leading term proportional to $y^\kappa$.
As remarked below (\ref{REMA}), these polynomials are the generalized
Jacobi polynomials $G_\kappa^{(a,b)}(y;\alpha)$.
The corresponding eigenvalue is
\begin{equation}
\prod_{j=1}^N \Big ( 1 + u \Big ( \kappa_j + (N - j)/\alpha + (a + b +1)/2 
 \Big )^2 \Big ).
\label{SEVJ}
\end{equation}

Now since each operator $\hat{D}_j^{(J)}$ is self-adjoint with respect to
the inner product (\ref{IPJ}), the operator (\ref{PEJ}) is also self-adjoint
with respect to this inner product. Furthermore, from the inequalities
(\ref{OR}) each eigenvalue (\ref{SEVJ}) is distinct. This immediately
implies (as has been proved before \cite[in the latter reference an
operator equivalent to (\ref{PEJ}) when restricted to acting on
symmetric functions of $y$ is also constructed]{beer93a,vandiej95c} 
that the generalized Jacobi polynomials
$\{G_\kappa^{(a,b)}(y;\alpha)
\}$ are orthogonal with respect to the inner product (\ref{IPJ}).
Note that the operator (\ref{PONS}) with $\hat{D}_j$ replaced by
$(\hat{D}_j^{(J)})^2$ can be used to show that $\{ E_\eta^{(J)} \}$ is
orthogonal with respect to (\ref{IPJ}).

\subsection{Generalized Jacobi polynomials with prescribed symmetry}
{}From the decomposition (5.2) and the fact that $(\hat{D}_j^{(J)})^2$
and $\tilde{H}^{(J,Ex)}$ both have unique polynomial eigenfunctions with
leading term $y^\eta$, we conclude that $E_{\eta}^{(J)}$ can be specified
as the eigenfunction of (\ref{HJC}) with leading term $y^\eta$. To pursue this
characterization, it is convenient to introduce the variable $y = \sin^2 \phi$
in (\ref{HJE}) and repeat the computation of (\ref{HTJE}). This gives
\begin{eqnarray}
- \tilde{H}^{(J,Ex)} & = & \sum_{j=1}^N y_j (1 - y_j) {\partial^2 \over
\partial y_j^2} + (a+1) \sum_{j=1}^N {\partial \over \partial y_j}
- (a+b + 2 +{2\over\alpha}(N-1))
\sum_{j=1}^N y_j{\partial \over \partial y_j} \nonumber
\\ & & + {1 \over \alpha}
\sum_{j \ne k}{1 \over y_j - y_k}
\Big ( 2y_j(1 - y_k) {\partial \over \partial y_j} -
{y_j(1 - y_k) \over y_j - y_k} (1 - M_{jk}) \Big )\nonumber \\&  := &U + V
\label{RAD}
\end{eqnarray}
where
\begin{equation}
U := -  \tilde{H}^{(C,Ex)} - (a+b+1+(N-1)/\alpha) \sum_{j=1}^N
y_j{\partial \over \partial y_j}, \quad
V =  \tilde{H}^{(L,Ex)} + \sum_{j=1}^N
y_j{\partial \over \partial y_j}.
\end{equation}
Following ref.~\cite{sogo96} we note that the operator $U$ is an eigenoperator
for the non-symmetric Jack polynomials $E_\eta^{(J)}$ while the operator
$V$ reduces by one the degree of a homogeneous polynomial (c.f.~(\ref{HTLE})).
Using these facts and noting $E_\eta(y,\alpha)$ has leading term $y^\eta$, we
see by proceeding as in the derivation of (\ref{EX1}) that
\begin{equation}
E_\eta^{(J)}(y,\alpha) = \Big (1 - (\hat{U} + V)^{-1}V \Big )E_\eta(y,\alpha).
\label{SO2}
\end{equation}
where with $\eta = P^{-1} \kappa$,
\begin{equation}
\hat{U} = U + \sum_{j=1}^N \Big ( \kappa_j^2 + 2 \kappa_j (N-j)/\alpha
+ \kappa_j (a+b+1) \Big ).
\end{equation}
Note that if this operator is expanded according to (\ref{OID}) the series
terminates after $|\eta|$ applications of $V$.

The generalized Jacobi polynomials with prescribed symmetry,
$S_{(\rho,\mu)}^{(J)}(y,\alpha)$ say, are defined as the eigenfunctions of 
(\ref{HJC}) given by the formula
\begin{equation}
S_{(\rho, \mu)}^{(J)}(y,\alpha) =\Big (1 - (\hat{U} + V)^{-1}V \Big )
S_{(\rho, \mu)}^{(J)}(y,\alpha).
\label{SO3}
\end{equation}
(for any polynomial $p$ which is an eigenfunction of $\hat{U}$, 
$(1 - (\hat{U} + V)^{-1}V)\,p$
is an eigenfunction of (\ref{HJC})). 
{}From (\ref{SO3}) and (\ref{SSUM}) it follows that 
\begin{equation}
S_{(\rho, \mu)}^{(J)}(y,\alpha) =
\sum_{\rm rearrangements} b_{Q^{-1}\kappa}
E_{Q^{-1}\kappa}^{(J)}(y,\alpha),
\end{equation}
which in turn implies (\ref{PCO}) is an eigenoperator for $\{S_{(\rho,
\mu)}^{(J)}
\}$ with $\hat{D}_j$ therein replaced by $(\hat{D}_j^{(J)})^2$. This operator is
self-adjoint with respect to  (\ref{IPJ}) and separates the eigenvalues, so
we conclude that $\{S_{(\rho, \mu)}^{(J)}\}$ is an orthogonal set 
with respect to the inner
product (\ref{IPJ}), with an appropriate change of variables (see
ref.~\cite[eq.~2.17]{forr96a}).

\subsection{Some special generalized Jacobi polynomials with 
prescribed symmetry}
Here we want to establish the analogue of (\ref{CON}) in the Jacobi case:
\begin{equation}
S_\eta^{(J)}(y,\alpha) = A_{\eta}^{(J)} \prod_{\gamma = 1}^p
\prod_{1 \le j < k \le N_\gamma} (y_k^{(\gamma)} - y_j^{(\gamma)}) \,
G_\rho^{(u,v)}(x_1,\dots,x_{N_0};p+\alpha),
\label{CONJJ}
\end{equation}
where $ A_{\eta}^{(J)}$ is some normalization, $\eta$ is given by 
(\ref{PK}), $\rho_1$ satisfies (\ref{IN}) and
\begin{equation}
u := {\alpha \over \alpha + p} \Big ( {p \over \alpha} + a + 1 \Big ) - 1,
\qquad v:=  {\alpha \over \alpha + p} \Big ( {p \over \alpha} + b + 1 \Big ) - 1
\end{equation}
Now set $N= N_0 + \sum_{\gamma=1}^p N_\gamma$ and denote the variables
$y_1,\dots,y_N$ as $\{x_j\}_{j=1,\dots,N_0}$ and $\{y_j^{(\gamma)}\}_{j=1,
\dots,N_\gamma}$ $(\gamma = 1,\dots,p)$.
Analogous to (\ref{DEC}), for $\tilde{H}^{(J,Ex)}$ acting on functions
symmetric in $\{x_j\}$ and anti-symmetric in $\{y_j^{(\gamma)} \}$ we make the
decomposition
\begin{equation}
\tilde{H}^{(J,Ex)} = \tilde{H}^{(J,x)} + \tilde{H}^{(J,y)} +\tilde{H}^{(J,xy)}
\end{equation}
where
\begin{eqnarray}
\tilde{H}^{(J,x)}   & = & \sum_{j=1}^{N_0} \Big (
x_j(1-x_j){\partial^2 \over
\partial x_j^2} + (a+1) {\partial \over \partial x_j} - (a+b+2+{2 \over 
\alpha}(N-1))x_j
{\partial \over \partial x_j} \Big )
\nonumber \\
& & +{2 \over \alpha} \sum_{j \ne k} {x_j (1 - x_k) \over x_j - x_k}
{\partial \over \partial x_j}
\end{eqnarray}
\begin{eqnarray}
\tilde{H}^{(J,xy)} & = & {1 \over \alpha}
 \sum_{\gamma=1}^p \sum_{j=1}^{N_\gamma} \sum_{k=1}^{N_0}
{1 \over y_j^{(\gamma)} - x_k} \left[ 
2 y_j^{(\gamma)} (1 - x_k)  {\partial \over \partial
y_j^{(\gamma)}} - 2 x_k (1 - y_j^{(\gamma)}) {\partial\over \partial
x_k} 
\nonumber \right.\\ && \left.
- { y_j^{(\gamma)} + x_k - 2y_j^{(\gamma)} x_k \over  y_j^{(\gamma)}
- x_k} (1 - M( y_j^{(\gamma)},x_k)) \right].
\end{eqnarray}
and $\tilde{H}^{(J,y)}$ is given by (\ref{HTJE}) with $y_1,y_2\dots,y_N$
replaced by $y_1^{(1)},y_2^{(1)},\dots,y_{N_p}^{(p)}$. In fact we have that
$\tilde{H}^{(J,y)}$ is a linear combination of $\tilde{H}^{(C,y)}$ and
$\tilde{H}^{(L,y)}$ and hence is an eigenoperator of $\prod_{\gamma = 1}^p
\Delta(y^{(\gamma)})$ with eigenvalue $\delta^{(J)}$ say. Also, from
(\ref{DEFHCWZ}) and (\ref{HLXY}) we have
$$
\tilde{H}^{(J,xy)} = - \tilde{H}^{(C,xy)} + \tilde{H}^{(L,xy)}
$$
so when acting on $F(x,y)$ as defined in Lemma 2.4 we see from 
Lemmas 2.4 and 4.5 that
$$
\tilde{H}^{(J,xy)} = {p \over \alpha} \sum_{j=1}^{N_0} x_j (1 - x_j)
{\partial^2 \over \partial x_j^2} + {2 \over \alpha} (N^{(z)} - p)
\sum_{j=1}^{N_0} x_j{\partial \over \partial x_j} + {p \over \alpha}
\sum_{j=1}^{N_0} {\partial \over \partial x_j}.
$$
Hence, for $\tilde{H}^{(J,Ex)}$ acting on $F(x,y)$ we have
$$
\tilde{H}^{(J,Ex)} = \delta^{(J)} + {p + \alpha \over \alpha} {\cal H}
$$
where ${\cal H}$ is the operator (\ref{RAD}) with $N=N_0$, the variables
$y_1, \dots,y_N$ replaced by $x_1,\dots,x_N$, $M_{jk} = 1$, $a=u$, $b=v$ and 
$\alpha$ replaced by $\alpha + p$. The operator ${\cal H}$ is the defining
eigenoperator for the generalized Jacobi polynomial in (\ref{CONJJ}),
thus establishing the validity of that equation.

In the
case $p=1$, $q=0$ we have
 the analogue of (\ref{ANTIS}) and thus another explicit formula:
\begin{equation}
S_\mu^{(J)}(y,\alpha) = \Delta (y) 
{1
\over v_{\kappa \kappa}(\alpha/(1+\alpha))}
G_\kappa^{(a,b)}(y;\alpha/(1+\alpha)).
\end{equation}
This result is derived from (\ref{HJE}) in the same manner as (\ref{ANTIS}) is
derived from (\ref{HCE}).

\section{Conclusion}
The Schr\"odinger operators (\ref{HCC}) and (\ref{HTHE})-(\ref{HTJE}) admit
polynomial eigenfunctions of the form given on the r.h.s.~of (\ref{nj}).
The most basic of these polynomials are the non-symmetric eigenfunctions of
(\ref{HCC}), which are referred to as the non-symmetric Jack polynomials.
In (\ref{EX2}), (\ref{EX5}), (\ref{SO2}) we give operator formulas which
transform the non-symmetric Jack polynomials to the non-symmetric
polynomial eigenfunctions of (\ref{HTHE})-(\ref{HTJE}). The operators
(\ref{HCC}) and (\ref{HTHE})-(\ref{HTJE}) also admit bases of fully symmetric
polynomial eigenfunctions, and polynomial eigenfunctions with a prescribed
symmetry. We have established orthogonality of these sets of eigenfunctions
with respect to inner products defined as multidimensional integrals, with
the corresponding symmetric ground state wave function as the weight function.
For the fully symmetric polynomials the orthogonality has previously been
established, but for the polynomials with prescribed symmetry this result
is new.

We expect the polynomials with prescribed symmetry to be relevant to the
calculation of correlation functions in the Calogero-Sutherland model with
spin (see e.g.~ref.~\cite{kk95} and references therein). For this purpose we
require normalization formulas, and an expansion
formula expressing the power  sum in terms of the Jack polynomials with
prescribed symmetry. Regarding the former, we point out that for the special
Jack polynomials with prescribed symmetry of Section 2.3, a conjecture for
the normalization has been made in ref.~\cite{forr96c}.

\vspace{.5cm}
\noindent
{\bf Acknowledgements}

\vspace{.2cm}
\noindent
We thank K.~Takemura (RIMS) for a useful remark, and T.~Miwa for his hospitality
at RIMS where this work was initiated. Also, correspondence with C.~Dunkl is
appreciated.

\bibliographystyle{unsrt}

\end{document}